\documentclass[pdftex]{pasj01}
\draft
\usepackage{natbib,color,ulem}
\usepackage{graphicx}

\Received{$\langle$reception date$\rangle$}
\Accepted{$\langle$acception date$\rangle$}
\Published{$\langle$publication date$\rangle$}

\begin{document}

\title{Nobeyama 45-m Mapping Observations toward Orion A. 
III. Multi-Line Observations toward  an Outflow-shocked Region, OMC-2 FIR 4}
\author{Fumitaka \textsc{Nakamura}\altaffilmark{1,2},
Shuri \textsc{Oyamada}\altaffilmark{1,3}, Sachiko Okumura\altaffilmark{3}, 
Shun \textsc{Ishii}\altaffilmark{1,4},
Yoshito \textsc{Shimajiri}\altaffilmark{5}, 
Yoshihiro \textsc{Tanabe}\altaffilmark{6}, 
Takashi \textsc{Tsukagoshi}\altaffilmark{6}, 
Ryohei \textsc{Kawabe}\altaffilmark{1,2}, Mumetake \textsc{Momose}\altaffilmark{6},
Yumiko \textsc{Urasawa}\altaffilmark{7}, 
Ryoichi \textsc{Nishi}\altaffilmark{7}, 
Sheng-Jun \textsc{Lin}\altaffilmark{8},
Shih-Ping \textsc{Lai}\altaffilmark{8},
Kazuhito \textsc{Dobashi}\altaffilmark{9},
Tomomi \textsc{Shimoikura}\altaffilmark{9},
Koji \textsc{Sugitani}\altaffilmark{10},
}%

\altaffiltext{1}{National Astronomical Observatory of Japan, 2-21-1 Osawa, Mitaka, Tokyo 181-8588, Japan}
\altaffiltext{2}{The Graduate University for Advanced Studies
(SOKENDAI), 2-21-1 Osawa, Mitaka, Tokyo 181-0015, Japan}
\altaffiltext{3}{Faculty of Science, Department of Mathematical and Physical Sciences, Japan Women's University, 
2-8-1 Mejirodai, Bunkyo-ku, Tokyo 112-8681, Japan}
\altaffiltext{4}{Joint ALMA Observatory, Alonso de C\' ordova 3107 Vitacura, Santiago, Chile}
\altaffiltext{5}{Laboratoire AIM, CEA/DSM-CNRS-Universit\' e Paris Diderot, IRFU/Service d'Astrophysique, CEA Saclay, F-91191 Gif-sur-Yvette, France}
\altaffiltext{6}{College of Science, Ibaraki University, 2-1-1 Bunkyo, Mito, Ibaraki 310-8512, Japan}
\altaffiltext{7}{Department of Physics, Niigata University, 8050 Ikarashi-2, Niigata 950-2181, Japan}
\altaffiltext{8}{ Institute of Astronomy and Department of Physics, National Tsing Hua University, Hsinchu 30013, Taiwan}
\altaffiltext{9}{Department of Astronomy and Earth Sciences, 
Tokyo Gakugei University, 4-1-1 Nukuikitamachi, Koganei, Tokyo 184-8501, Japan}
\altaffiltext{10}{Graduate School of Natural Sciences, Nagoya City University, Mizuho-ku, Nagoya, Aichi 467-8501, Japan}


\maketitle

\begin{abstract}
We present the results of mapping observations toward an outflow-shocked region, OMC-2 FIR 4
using the Nobeyama 45-m telescope.
We observed the area in $^{13}$CO ($J=1-0$), C$^{18}$O ($J=1-0$), N$_2$H$^+$ ($J=1-0$), 
CCS ($J_N=8_7-7_6$), HCO$^+$ ($J=1-0$), H$^{13}$CO$^+$ ($J=1-0$), 
HN$^{13}$C ($J=1-0$), H$^{13}$CN ($J=1-0$), DNC ($J=1-0$), N$_2$D$^+$ ($J=1-0$), and
DC$_3$N  ($J=9-8$).  We detected a dense molecular clump that contains FIR 4/5.
We also detected in $^{13}$CO  blueshifted and redshifted components driven presumably by protostellar outflows in this region.
The axes of the FIR 3 and VLA 13 outflows, projected on the plane of the sky, appear to point toward the FIR 4 clump, suggesting
 that the clump may be compressed by protostellar outflows from Class I sources, FIR 3 and VLA 13.
Applying the hyperfine fit of N$_2$H$^+$ lines, 
we estimated the excitation temperature to be $\sim$ 20 K. 
The high excitation temperature 
is consistent with the fact that the clump contains protostars. 
The CCS emission was detected in this region for the first time.  Its abundance is estimated to be
a few $\times 10^{-12}$, indicating that the region is chemically evolved at $\sim 10^5$ years, which
is comparable to the typical lifetime of the Class I protostars.
This timescale is consistent with the scenario that star formation in FIR 4 is triggered by
dynamical compression of the protostellar outflows.
The [HNC]/[HCN] ratio is evaluated to be $\sim 0.5$ in the dense clump and the outflow lobes, 
whereas it is somewhat larger in the envelope of the dense clump. 
The small [HNC]/[HCN] ratio indicates that the HNC formation was prevented due to high temperatures.  
Such high temperatures seem to be consistent with the scenario that either protostellar radiation or outflow compression, or both, 
affected the thermal properties of this region.
\end{abstract}

\section{Introduction}
\label{sec:intro}

In cluster-forming regions, stellar feedback from forming stars often
influences the surrounding environments \citep{mckee07,wang10,krumholz14}. 
For example, protostellar outflows can dynamically compress the 
surrounding  gas and/or  adjacent dense cores, 
and promote fragmentation and gravitational contraction 
leading to star formation \citep{shimajiri08}.
In addition, such stellar feedback injects significant momentum
in molecular clouds, controlling the efficiency of star formation
\citep{mckee89, nakamura07, nakamura12, nakamura14, arce11}.
Theoretically, the negative effect of the stellar feedback is more important than the positive effect
\citep{li06,nakamura07,krumholz14}, but the positive effect is sometimes observed to 
enhance local star formation \citep{elmegreen77,sugitani94,chauhan09}.
Among the stellar feedback processes, the protostellar outflow feedback
can trigger local star formation through dynamical compression of dense parts
in molecular clouds \citep{sandell01,yokogawa03,shimajiri08}.

One of the good examples of the dynamical interaction between outflow and dense gas  is
the FIR 4 region  in the Orion Molecular Could 2 (OMC-2),  
in which fragmentation seems to be induced by the outflow shock from a nearby protostar FIR 3 \citep{shimajiri08, lopez13}.
Multi-epoch, multiwavelength Very Large Array (VLA) observations 
 \citep{osorio17} support that the FIR 3 outflow is actually moving toward the FIR 4 region, suggesting that
 the FIR 3 outflow is interacting with the FIR 4 region \citep[see also][]{gonzalez16}.
 These morphological evidences strongly support that the outflow is interacting with the dense gas  in this region.
\citet{shimajiri08} detected 11 cores on the basis of the interferometric 
3-mm dust continuum observations [see also \citet{hacar18} for the Atacama Large Millimeter/submillimeter Array (ALMA) observations].
Some of the cores have masses greater than the thermal Jeans mass in this region.  
Near-infrared observations suggest that protostars are already 
created in this clump.  Although the luminosity of the central protostar is not so prominent, 
the mass of the protostellar envelope is derived 
to be about 27 $M_\odot$ \citep{furlan14}.  
Therefore, this protostar is likely to  evolve at least into an intermediate-mass star.  
Since there are several fragments with masses greater 
than the local Jeans mass, the FIR 4 region may be forming a small star cluster
consisting of low-mass and intermediate-mass stars.

Spectral line surveys toward FIR 4 also suggest that FIR 4 is likely to be a hot core \citep{ceccarelli10,kama13}.
The hot core might be formed by the outflow compression. Or, it might be created before the outflow compression which promotes
fragmentation in the FIR 4 clump. In contrast,
very recently, \citet{favre18} found that this region has relatively uniform excitation temperature of c-C$_2$H$_3$
which led them to claim that there is no clear evidence of the outflow compression since the compression  should create
the temperature gradient.  
They proposed that the FIR 3 outflow does not hit FIR 4 and the spatial overlap
of the outflow lobe and the clump in the two-dimensional map may be the apparent coincidence 
along the line of sight.
However, because of the large uncertainty in the estimated temperatures
and poor angular resolution, it  would be difficult to accurately discuss the temperature distribution in this region from the data. 
Higher sensitivity and higher angular resolution observations are needed to further constrain the outflow compression scenario.

Located in the Orion A molecular cloud, the FIR 4 region is one of the closest cluster-forming regions.
The distance to the Orion A molecular cloud is estimated to be 414 $\pm7$ pc by \citet{menten07}  
 and 418 $\pm$ 6 pc by \citet{kim08}.
These two measurements reasonably coincide with each other.
Hereafter, we adopt 414 pc as a representative distance of the FIR 4 region.
Some indirect evidence of dynamical interaction by the FIR 3 outflow also comes from recent spectral line surveys \citep{kama13, shimajiri15b, kama15}.
In the present paper, we characterize the physical and chemical properties of the possible outflow-shocked region, FIR 4,
by several molecular line observations such as N$_2$H$^+$, CCS, HN$^{13}$C, and H$^{13}$CN.
In addition, we identify possible protostellar outflow candidates in this region, using the $^{13}$CO ($J=1-0$) and HCO$^+$ ($J=1-0$) data.

The observations of the FIR 4 region  were made part of the Nobeyama Star Formation Legacy Project at the Nobeyama Radio Observatory (NRO) to
observe nearby star-forming regions such as Orion A, Aquila Rift, and M17. A overview of the project will be presented in a separate paper
(Nakamura et al. 2019) and the detailed observational results for the individual regions are given in other
articles (Orion A: Tanabe et al. 2019, \citet{ishii19}, 
Aquila Rift: \citet{shimoikura19}, \citet{kusune19},
M17: \citet{shimoikura19}, \citet{sugitani19}
other regions: \citet{dobashi18}).  The FIR 4 region was observed once or twice every day 
to check the absolute flux calibration of $^{13}$CO ($J=1-0$), C$^{18}$O ($J=1-0$), and 
N$_2$H$^+$ ($J=1-0$) in Orion A, which were obtained with a new receiver, FOREST
installed on the Nobeyama 45-m telescope.
Thus, these molecular line data have excellent sensitivity.

The paper is organized as follows. Section \ref{sec:obs} describes
the detail of our Nobeyama 45-m observations. 
In Section \ref{sec:results}, we present the results of our mapping observations.  
We report the detection of the CCS emission at 93 GHz, which is the first detection of CCS in the OMC-2 region.  
In Section \ref{sec:derivation}, we discuss some characteristics of the internal structure of this region.
Using our molecular line data, we identify the molecular outflow candidates in this region in Section \ref{sec:outflow}, 
and briefly discuss how the outflows interact with the dense clump.
Finally, Section \ref{sec:summary} summarizes the main results 
of this study.

\section{Observations and Data}
\label{sec:obs}

\begin{table}
  \tbl{Observed lines with FOREST}{%
  \begin{tabular}{lllll}
  \hline
  Molecule &Transition & Rest Frequency & Effective Resolution$^b$
 & noise level$^c$ \\
& &(GHz)  & (arcmin) & (K) \\ 
\hline
$^{13}$CO	& $J$=1--0 & 	110.201354	& \timeform{21.7''}	 & 0.19   \\
C$^{18}$O	& $J$=1--0 &	109.782176 	& \timeform{21.7''}  & 0.16  \\
CCS	& $J_N=8_7-7_6$ &	93.870098 	& \timeform{32.5''} 	& 0.04 \\
N$_2$H$^+$	& $J=1-0, F_1F \rightarrow F_1'F' =23\rightarrow 12$ &  93.1737637$^a$ 	& \timeform{23.4''}  & 0.06  \\
\hline
  \end{tabular}}\label{tab:forest}
  \begin{tabnote}
 $^a$the rest frequency of the main hyperfine component from \citet{pagani09}.
In the last column, we present the rms noise level in $T_{\rm mb}$. 
We observed all 6 hyperfine components.

$^b$The CCS data were spatially smoothed to improve the signal-to-noise ratio. 

$^c$ The noise levels are measured with a velocity resolution of 0.05 km s$^{-1}$.
  \end{tabnote}
\end{table}

\subsection{$^{13}$CO 
($J$ = 1--0), C$^{18}$O ($J$ = 1--0), N$_2$H$^+$ ($J$=1--0) and , CCS $(J_N=8_7-7_6$) Observations}

\begin{figure}
\begin{center}
\includegraphics[width=0.8 \textwidth,bb=0 0 471 395]{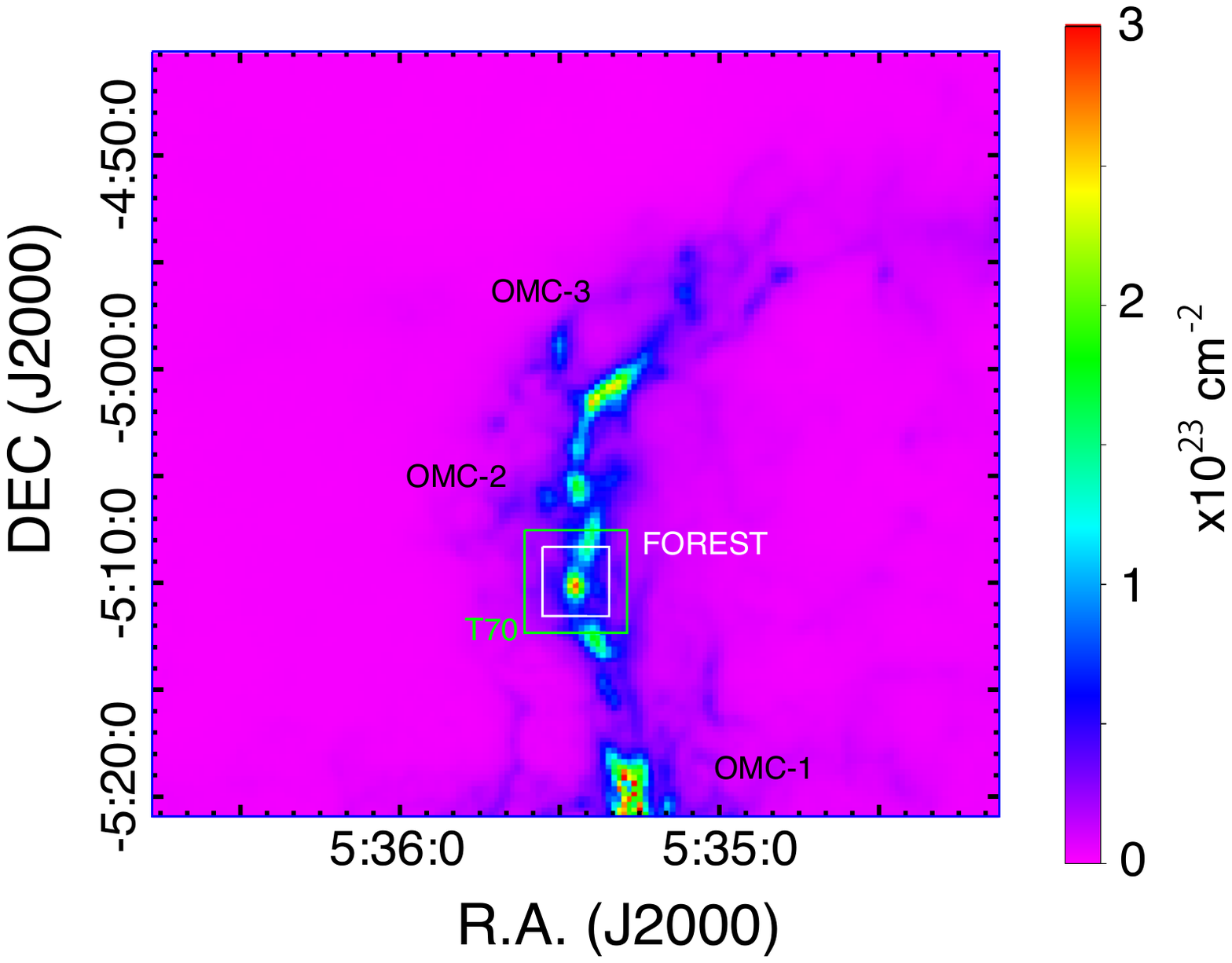}
\caption{Mapped areas in Orion A.  The color image shows the {\it Planck $-$ Herschel} H$_2$ column density created by \citet{lombardi14}.  
The areas mapped with FOREST and T70 are indicated in a white and green squares, respectively.  We also indicated the regions of OMC-1, OMC-2, and OMC-3.}
\label{fig:obsbox}
\end{center}
\end{figure}

We carried out mapping observations in $^{13}$CO 
($J$ = 1--0), C$^{18}$O ($J$ = 1--0), N$_2$H$^+$ ($J$=1--0) and CCS $(J_N=8_7-7_6$) toward the FIR3/4/5 region in Orion A 
using a new receiver FOREST [Four-Beam Receiver System on the 45-m Telescope] 
which is  a 4-beam, dual-polarization sideband-separating SIS receiver
installed on the 45 m telescope of the Nobeyama Radio Observatory. 
See \citet{minamidani16} for more details of FOREST.
The telescope has an Full-Width at Half Maximum (FWHM) beam size of \timeform{18"} at 93 GHz.
The beam separation of FOREST is $\timeform{51.7"}$ on the sky.  
The mapping area is indicated in Figure \ref{fig:obsbox} with a white square.
The observed lines and their parameters are summarized in Table \ref{tab:forest}.
The observations were done in a On-The-Fly (OTF) mode 
\citep{sawada08}
in the period from 2016 December to 2017 March.  
We adopted a spectral window mode which allows us to obtain 4 lines simultaneously.
The FIR 4 maps were obtained for the intensity calibration of the larger Orion A map
(see Nakamura et al. 2018 for more details).
As the backends, we used a digital spectrometer based on an FX-type correlator, SAM45, that is 16 sets of 4096 channel array.
The frequency resolution of all spectrometer arrays was set to 15.26 kHz, which corresponds to 0.05 km s$^{-1}$ at 93 GHz.
The scan interval of the OTF observation is set to \timeform{5''.17}, so that individual scans by the 4 beams of FOREST are fully overlapped.

The temperature scale was determined by the chopper-wheel method.
The telescope pointing was checked every 1 hour by observing the SiO maser line from Orion KL (R.A. [J2000], Dec. [J2000]) = 
$\timeform{5h35m14.5s}, \timeform{-5D22'30.4''}$). The pointing accuracy was better than $\sim$ 3$''$ throughout the entire observation.
The typical system noise temperature was in the range from 150 K to 200 K  in the single sideband mode at the observed elevation of El=\timeform{30D}--\timeform{50D}.

In order to minimize the scanning effects, the data with orthogonal scanning directions along the R.A. 
($x$-scan) and Dec. ($y$-scan) axes were combined into a single map.
In total, we combined 6 $x$-scan and 7 $y$-scan data into single maps.
We adopted a spheroidal function with a width of 7$''$.5 as a gridding convolution function to calculate the intensity at each grid point 
of the final cube data with a spatial grid size of \timeform{7.5"}, 
about a third of the beam size. 
The resultant effective angular resolution was about 22$''$ at 110 GHz, 
corresponding to $\sim$ 0.05 pc at a distance of 414 pc. 

The main beam efficiencies were 0.50 and 0.43 at 93 GHz and 110 GHz, respectively.
We divided the intensities by the main beam efficiencies at the corresponding frequencies and made the maps in the brightness temperature scale. 
More details of the observations and data reduction will be given in a separate paper
(Nakamura et al. 2018).
The CCS image presented in figure 2(d) was smoothed to improve the signal-to-noise
ratios of the map since the CCS emission in the FIR 4 region is very weak.

\begin{table}
  \tbl{Observed lines with T70}{%
  \begin{tabular}{lllll}
  \hline
  Molecule &Transition & Rest Frequency &Effective Resolution$^a$
& noise level \\
& &(GHz)  & (arcmin) & (K) \\ 
\hline
HCO$^+$	& $J$=1--0 & 	 89.188526 	& \timeform{24.9''}	& 0.45  \\
H$^{13}$CO$^+$	& $J$=1--0 & 	86.754288	& \timeform{33.6''}	&0.20  \\
H$^{13}$CN	& $J$=1--0, $F$=2--1 &	86.3401764 	& \timeform{33.6''}  & 0.16 \\
HN$^{13}$C	& $J$=1--0, $F$=2--1 &	 87.090859  	& \timeform{33.6''} & 0.18 \\
N$_2$D$^+$	& $J$=1--0 &  77.107798 	& \timeform{32.5''} & 0.29 \\
DNC	& $J$=1--0 &  76.305697 	& \timeform{32.5''} & 0.27 \\
DC$_3$N	& $J$=9--8 &  75.987149 	& \timeform{32.5''} & 0.33 \\
\hline
  \end{tabular}}\label{tab:t70}
  \begin{tabnote}
 $^a$ For all the lines except HCO$^+$, the images were smoothed to improve the signal-to-noise 
 ratios.
The last column is average rms noise levels of the whole area at an velocity resolution of 0.05 km s$^{-1}$. 
  \end{tabnote}
\end{table}

\subsection{HCO$^+$ ($J$ = 1--0), H$^{13}$CO$^+$ ($J$ = 1--0), HN$^{13}$C ($J$ = 1--0), H$^{13}$CN ($J$ = 1--0), 
DNC ($J$ = 1--0), N$_2$D$^+$ ($J=1-0$), and DC$_3$N ($J$ = 9--8) Observations}

We carried out mapping observations in 
HCO$^+$ ($J$ = 1--0), H$^{13}$CO$^+$,  
($J$ = 1--0), H$^{13}$CN ($J$ = 1--0), HN$^{13}$C ($J$ = 1--0), 
DNC ($J$ = 1--0), N$_2$D$^+$ ($J$ =1--0), and DC$_3$N ($J$=9--8) 
toward slightly larger area than we mapped with FOREST,
using the T70 receiver installed on the Nobeyama 45-m telescope.
The mapping area is indicated in Figure \ref{fig:obsbox} with a green square.
The observed lines and their parameters are summarized in Table \ref{tab:t70}.
The observation procedure was essentially the same as that of the FOREST observations.
The observations were done in the OTF mode in the period from 2016 December to 2017 February. 
We used SAM 45 at a frequency resolution of 3.81 kHz, corresponding to
$\sim$ 0.0125 km s$^{-1}$ at $\sim$ 86 GHz.
Scan interval of the OTF observation is set to \timeform{5''}. 
The typical system noise temperature was in the range from 150 K to 250 K 
in the single sideband mode at the observed elevation.

The temperature scale was determined by the chopper-wheel method.
The telescope pointing was checked every 1 hour by observing the SiO maser line from Orion KL.
The pointing accuracy was better than $\sim$ 3$''$ throughout the entire observation.
Adopting a spheroidal function with a width of 7.5$''$ 
as a gridding convolution function to compute the intensity at each grid point, 
we made final images with a spatial grid size of \timeform{7.5"} by 
 combining a $x$-scan and a $y$-scan images into single maps to 
obtain a map in the antenna temperature scale. 
The resultant effective angular resolution was about 25$''$.
Finally, we divided the intensities by  
the main beam efficiencies at the corresponding frequencies 
to obtain the maps in the brightness
temperature scale.
The main beam efficiencies at 89 GHz, 87 GHz and 76GHz for the T70 receiver were about 0.543, 0.544, and 0.549, respectively.
All the lines except HCO$^+$ are relatively weak. To improve the signal-to-noise
ratios of the maps, 
the images presented below for all the observed lines except HCO$^+$  were smoothed
and thus the resultant effective angular resolutions were $\sim$ 33$\arcsec$
for those images.


%

\section{Spatial Distributions of Molecular Line Emission}
\label{sec:results}

In this section, we describe the spatial distributoins of
 the molecular line emission in the observed area.
In Figure \ref{fig:co}, we present the integrated intensity maps of the molecular line emission detected.
Recently, \citet{kainulainen17} investigated the OMC-2 region with ALMA.
Following their Table 2, the protostars previously identified in this region are listed 
in Table \ref{tab:protostars}.

\begin{table}
  \tbl{Protostars in the FIR 3/4/5 Region}{%
  \begin{tabular}{lllllllll}
  \hline
Name  & R.A. [J2000] & Dec. [J2000] & Mass & Class & Outflow candidate$^a$ & Identification$^b$ & Note & References$^c$ \\ \hline
HOPS68 & 05:35:24.287 & $-$05:08:30.65 & 1.7 & I  & R & C & FIR2 & \citet{aso00,williams03} \\
  & & & & & & & & \citet{takahashi08}\\ 
HOPS66 & 05:35:26.928 & $-$05:09:24.40 & 1.5 & Flat & &  & &\citet{takahashi08}  detected near-infrared knots. \\
HOPS370 & 05:35:27.618 & $-$05:09:34.06 & 2.5 & I & BR & C & FIR3 &\citet{aso00,williams03} \\
  & & & & & & & & \citet{wu05}  \\    
    & & & & & & & & \citet{takahashi08,shimajiri08}  \\    
HOPS65 & 05:35:21.566 & $-$05:09:38.62 & 0.2 & I & &  & & \\ 
HOPS64 & 05:35:26.928 & $-$05:09:54.51 & 0.5 & ? & R & M &  & this paper \\
HOPS108 & 05:35:27.015 & $-$05:09:59.59 & 0.3 & 0 & BR & M &  FIR4 & \citet{takahashi08} claimed marginal detection \\
HOPS368 & 05:35:24.717 & $-$05:10:29.78 & 0.7 & I  & BR & C & VLA 13 & \citet{takahashi08} \\ \hline
\hline
  \end{tabular}}\label{tab:protostars}
  \begin{tabnote}
 $^a$ Outflow identification is based on the present paper. See Section \ref{sec:outflow} for details.
 
 $^b$ C and M indicate outflows identified as clear and marginal, respectively.
 
 $^c$ previous detection

  \end{tabnote}
\end{table}

\subsection{$^{13}$CO and C$^{18}$O}

\begin{figure}
\begin{center}
\includegraphics[width=0.25 \textwidth, bb=0 -50 476 420]{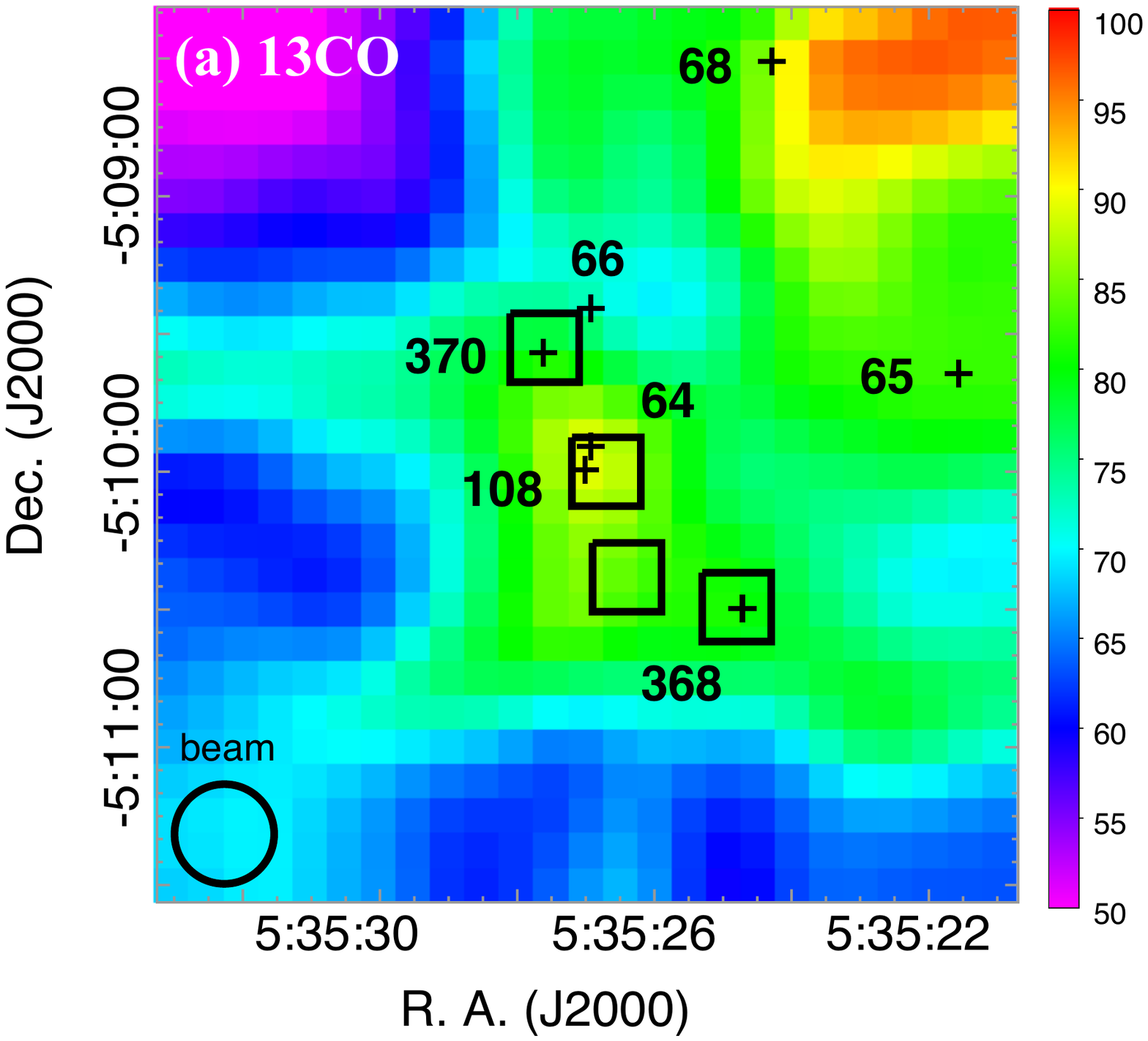}
\includegraphics[width=0.25 \textwidth,bb=0 -50 473 419]{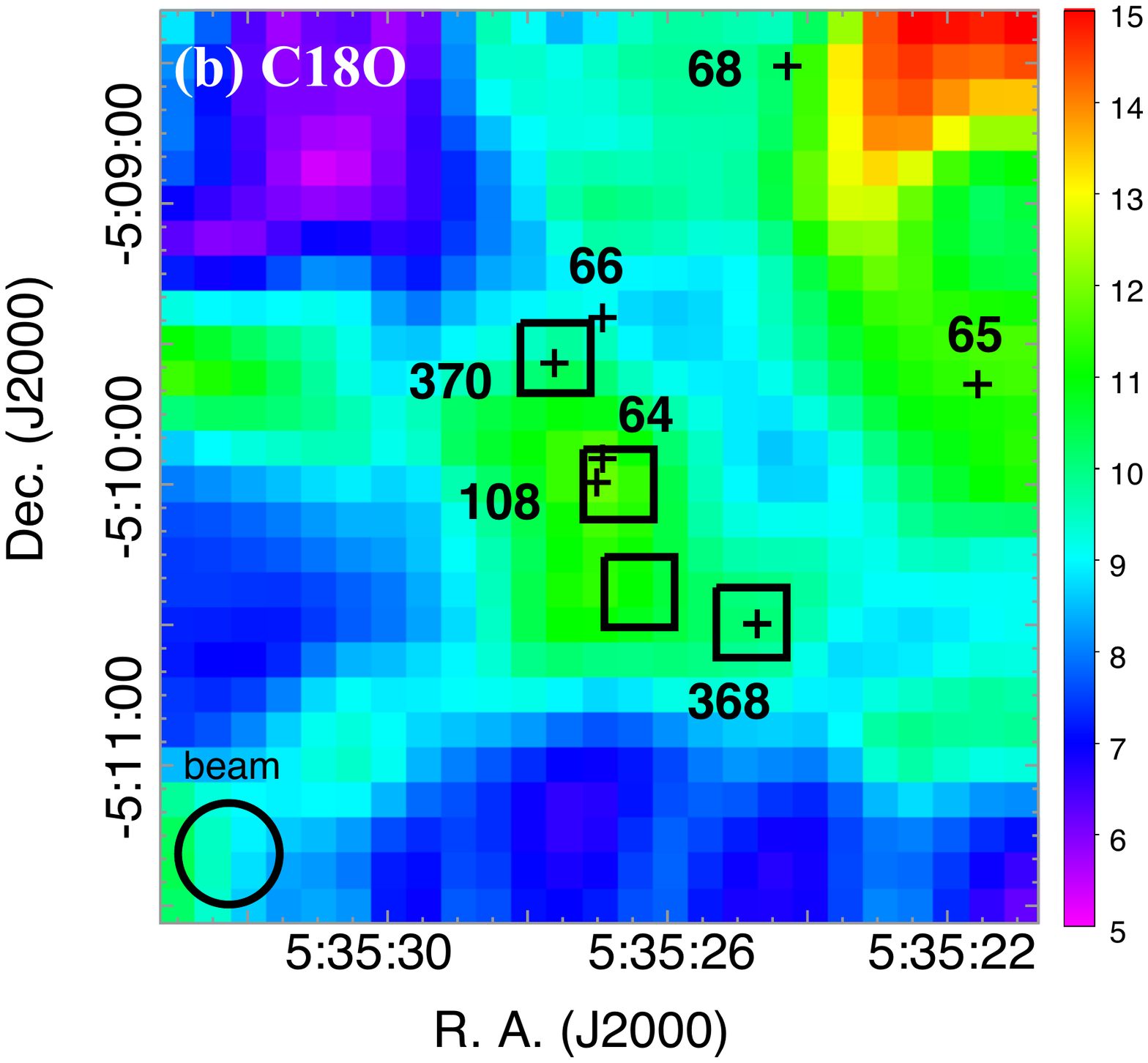}
\includegraphics[width=0.25 \textwidth, bb=0 -50 476 422]{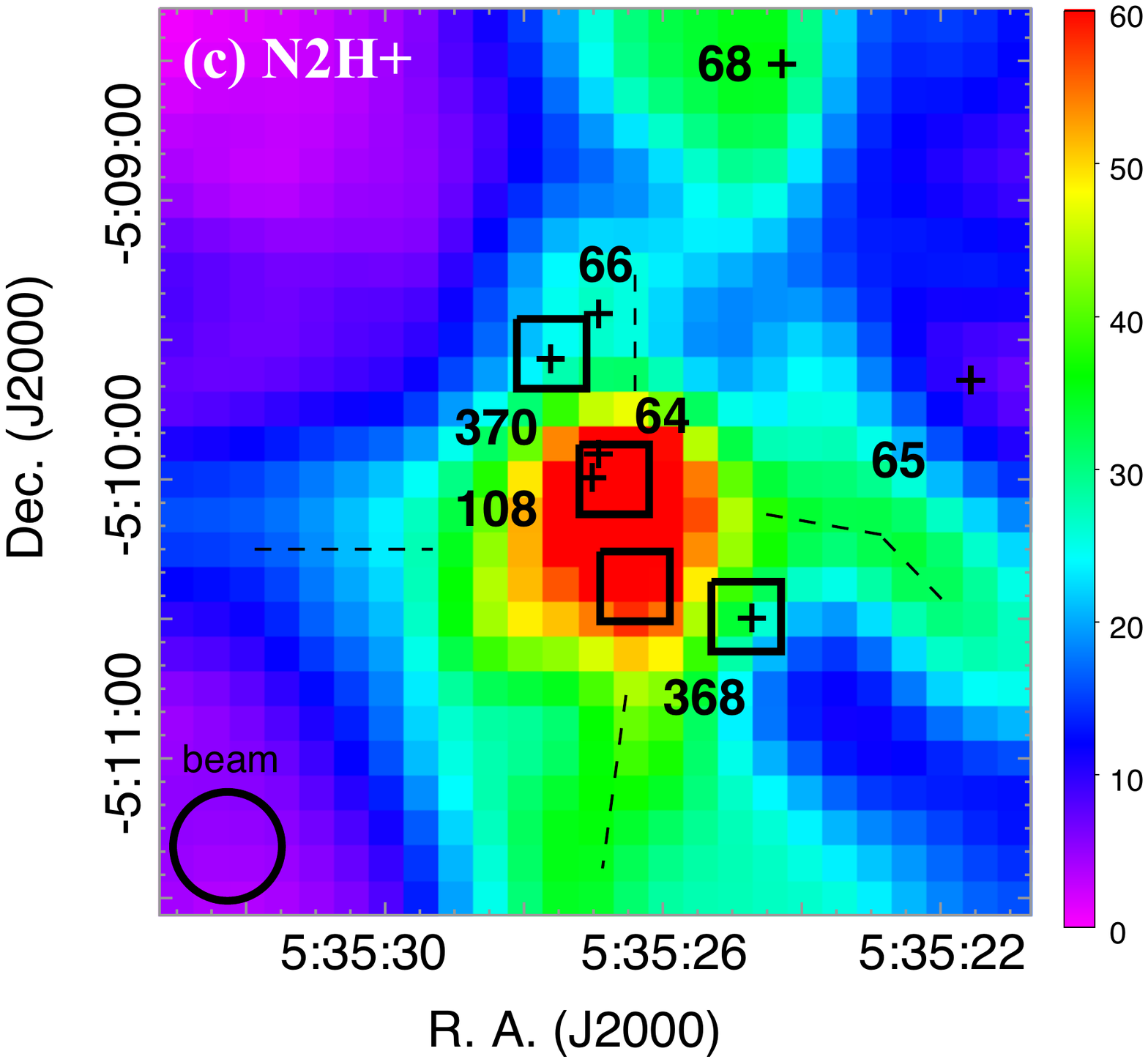}
\includegraphics[width=0.25 \textwidth, bb=0 -50 475 500]{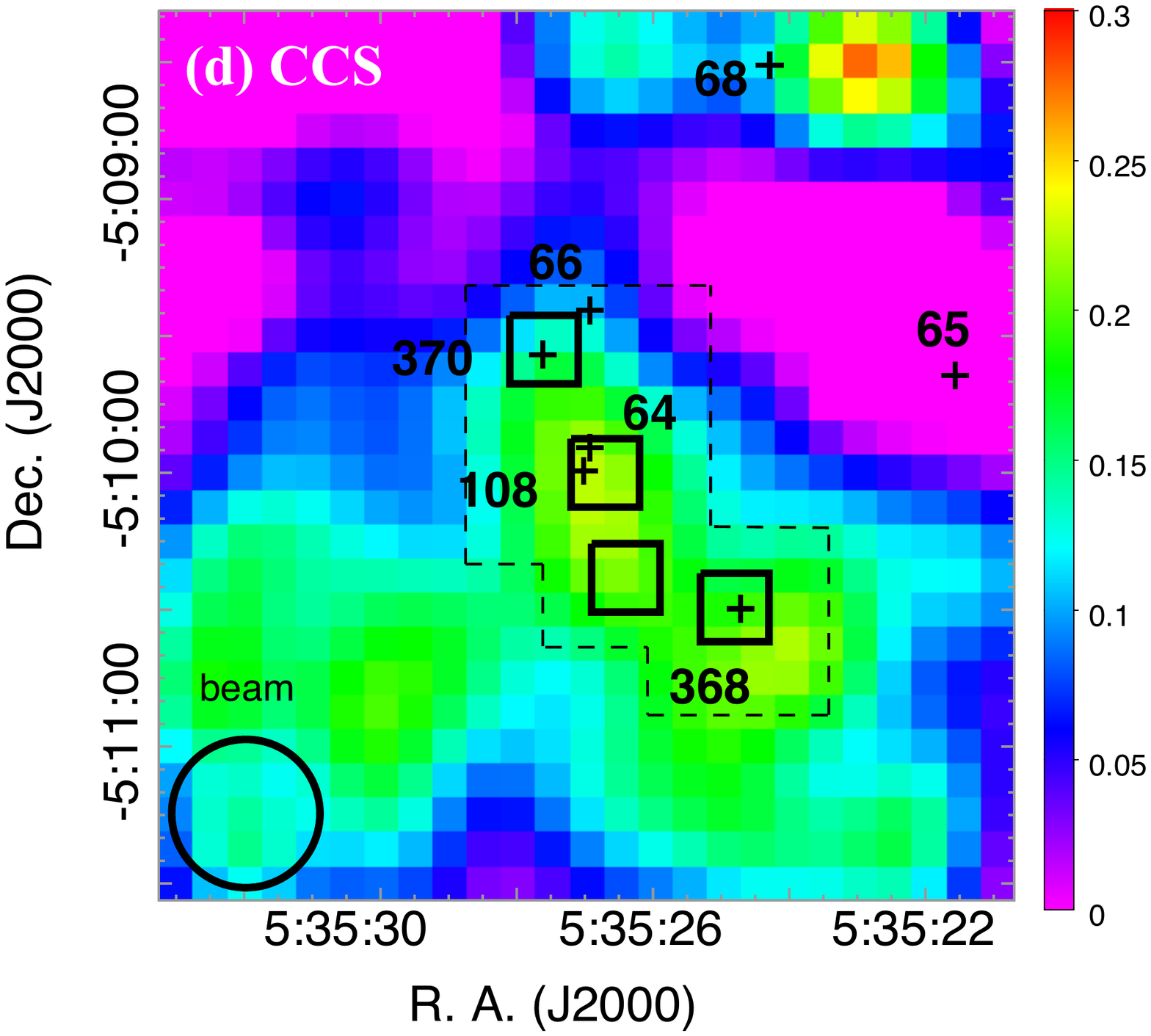}
\includegraphics[width=0.25 \textwidth, bb=0 -50 472 500]{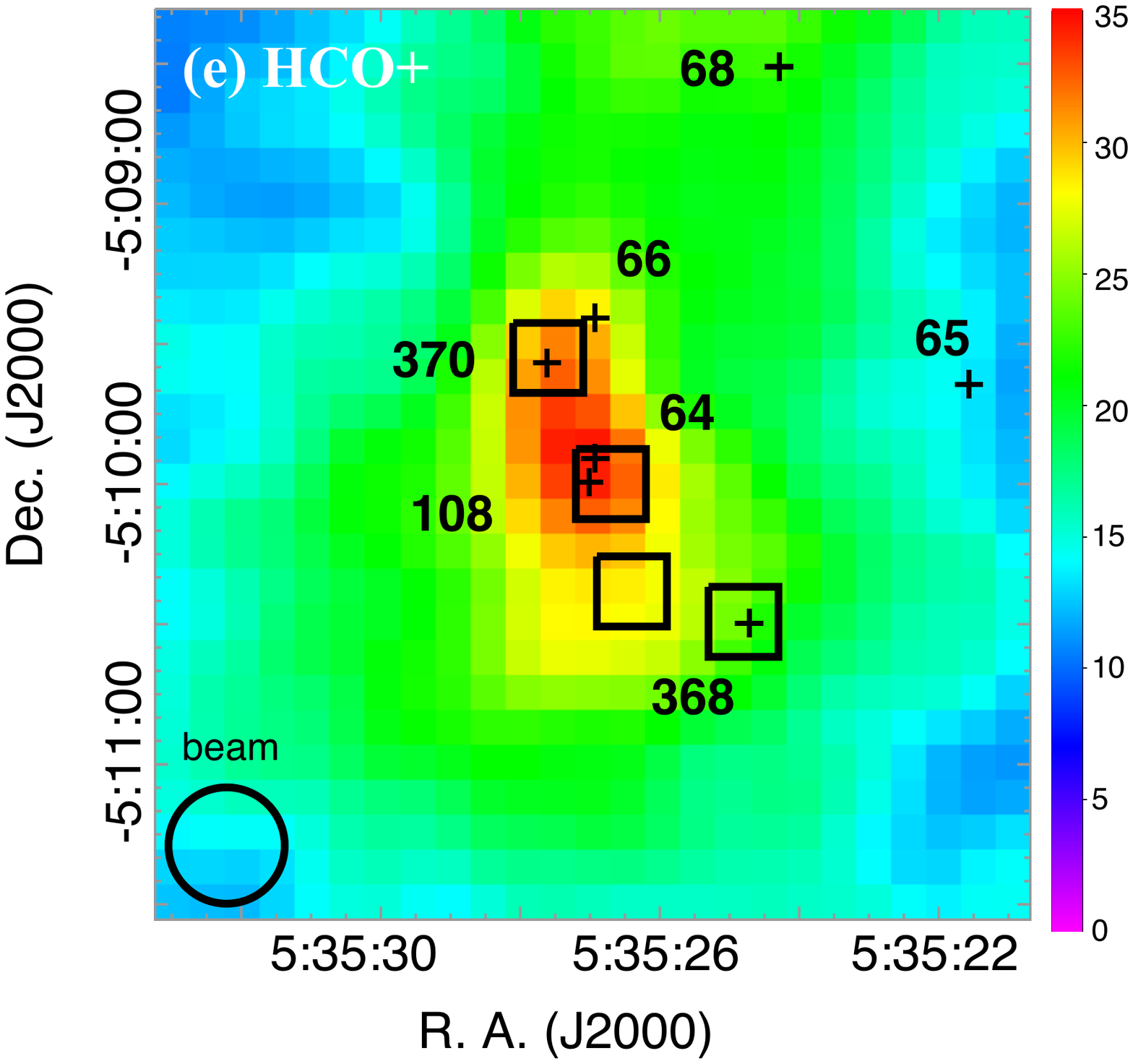}
\includegraphics[width=0.25 \textwidth, bb=-100 -200 300 350]{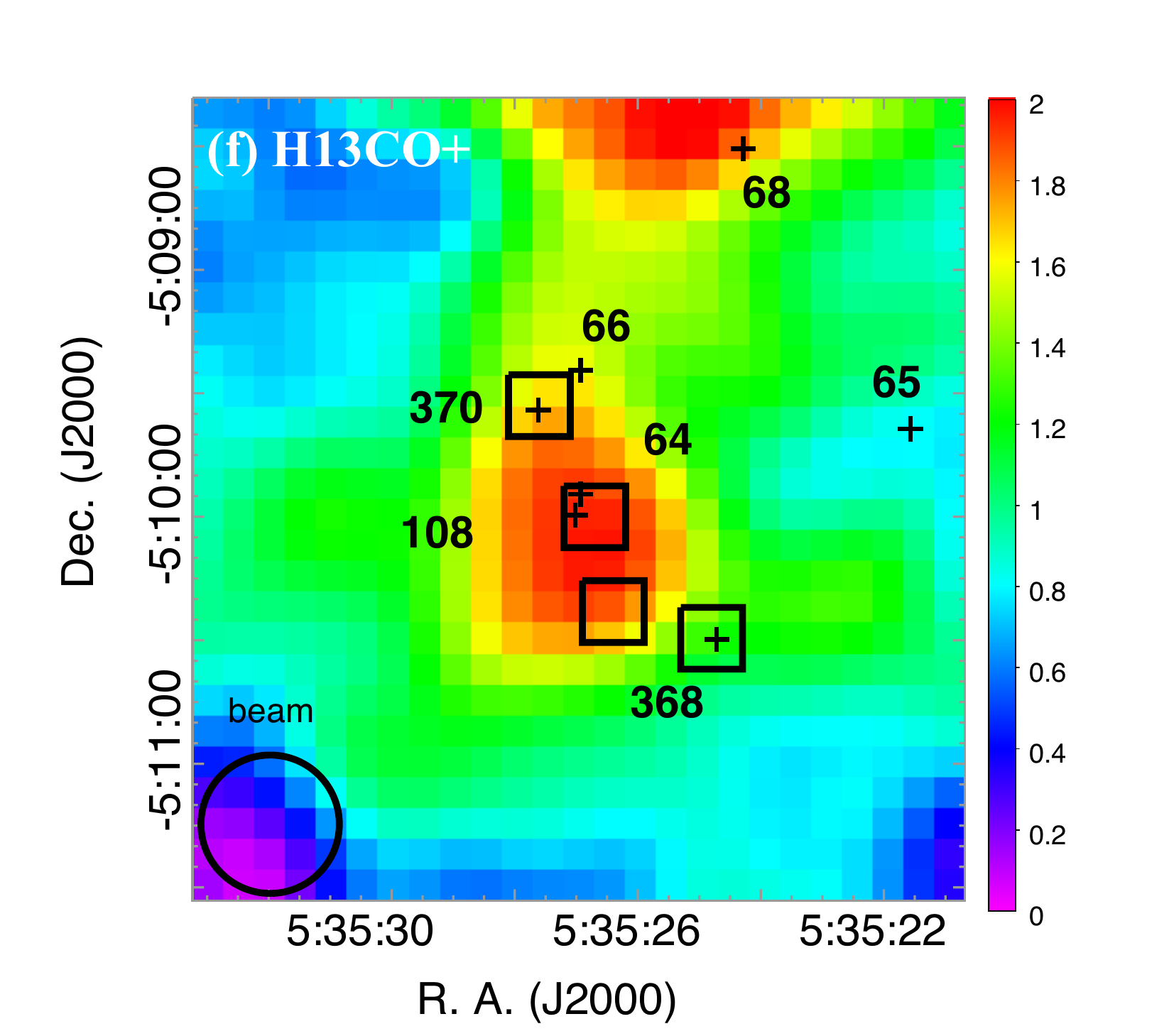}
\includegraphics[width=0.25 \textwidth, bb= 0 -150 473 470]{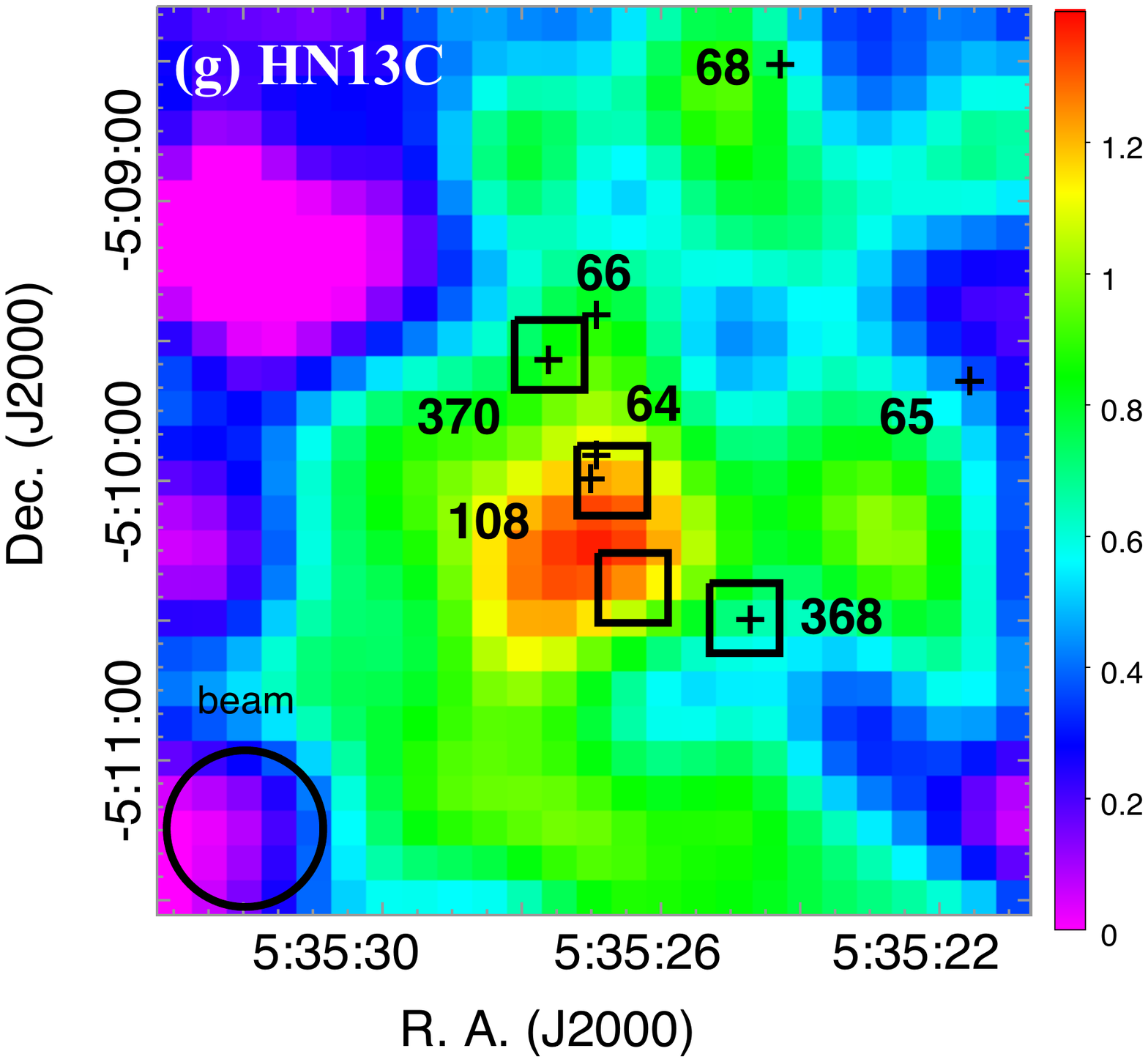}
\includegraphics[width=0.25 \textwidth, bb=0 -150 475 470]{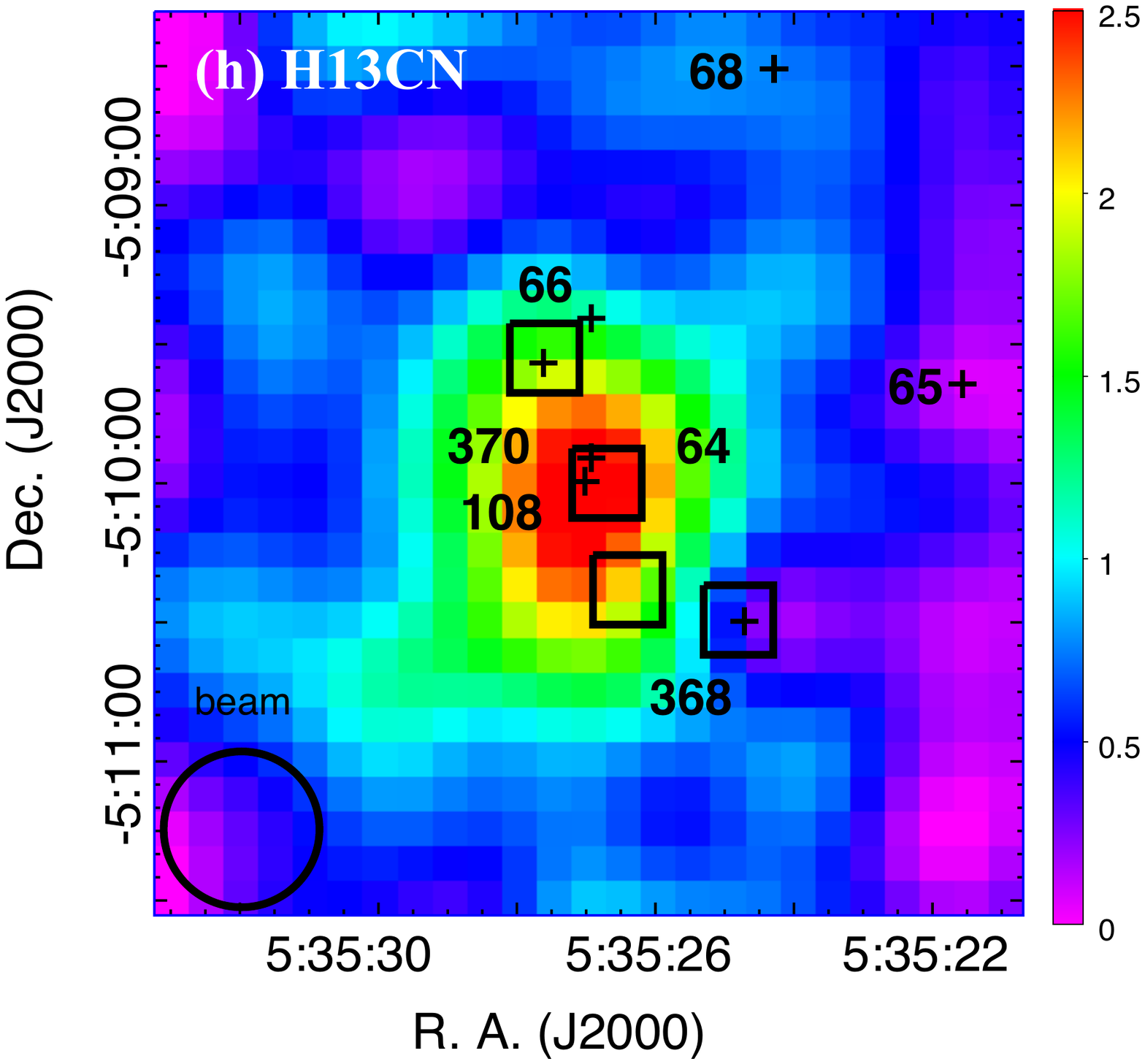}
\includegraphics[width=0.25 \textwidth, bb=-50 -250 350 400]{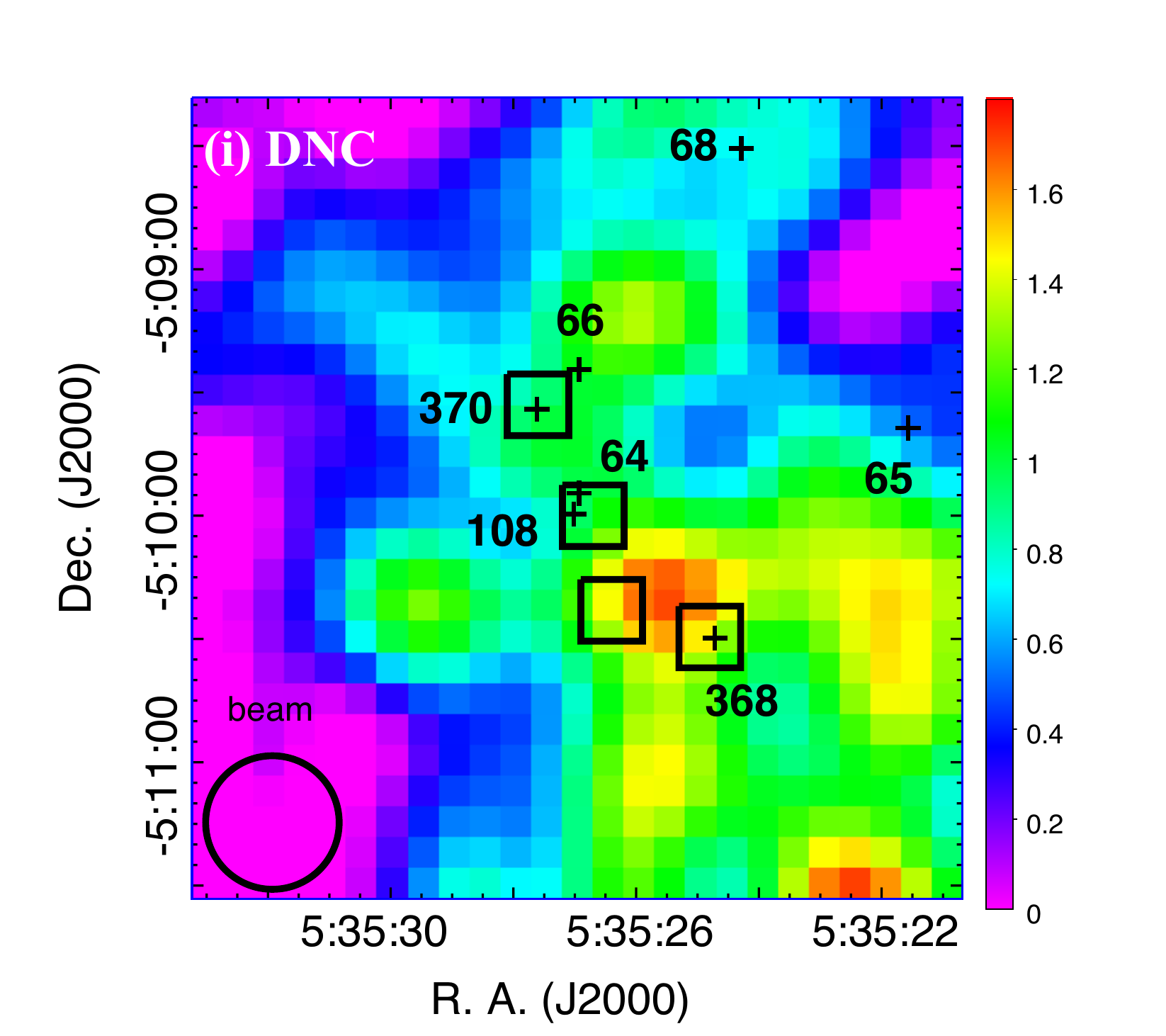}
\end{center}
\caption{Total integrated intensity maps of the FIR 4 region.  (a) $^{13}$CO ($J=1-0$), (b) C$^{18}$O  ($J=1-0$),
(c) N$_2$H$^+$ ($J=1-0$), (d) CCS ($J_N=8_7-7_6$), (e) HCO$^+$  ($J=1-0$),
(f) H$^{13}$CO$^+$ ($J=1-0$), (g) HN$^{13}$C  ($J=1-0$), (h) H$^{13}$CN ($J=1-0$), and (i)
DNC  ($J=1-0$) . 
The integrated intensities are given in K km s$^{-1}$ in the brightness temperature scale.
The squares from north to south indicate FIR 3, FIR4, FIR5 dust cores  identified by \citet{chini97}, and VLA13 protostar \citep{reipurth99}.
The positions of the protostars listed in Table \ref{tab:protostars} are shown with the crosses.
The numbers in the panels denote HOPS ID of the protostars.
The effective angular resolution of the maps are  shown in the bottom-left of each panel.
}
\label{fig:co}
\end{figure}

In Figures \ref{fig:co}a and \ref{fig:co}b, we present the velocity-integrated intensity maps of $^{13}$CO and C$^{18}$O, respectively.
The velocity range of the integration is set to 5.1 km s$^{-1}$ $-$ 15.1 km s$^{-1}$
for both $^{13}$CO and C$^{18}$O.
The distribution of the $^{13}$CO emission matches reasonably well with that of C$^{18}$O.
For both $^{13}$CO and C$^{18}$O maps, the intensities are strongest near the upper-right corner of the observed area
(recognized as parts with red in Figures \ref{fig:co}a and \ref{fig:co}b), 
which is a part of the other core traced in the 1.1-mm continuum 
 [Core No. 320 of \citet{shimajiri15}, see also the white box in Figure \ref{fig:obsbox}.].
The $^{13}$CO emission takes its local maximum near the position of FIR 4 (HOPS108)
at (R.A. [J2000], Dec. [J2000]) = (\timeform{5h35m27s}, \timeform{-5D10'00"}), 
which coincides with the position of the C$^{18}$O local peak. 
Both the $^{13}$CO and C$^{18}$O emission trace a dense clump and its envelope which contains
FIR 3/4/5.  The dense clump corresponds to Core No. 327 of \citet{shimajiri15}
and is elongated in the north-south direction, which is
roughly similar to the large-scale filament axis of the OMC-2 region.
The high resolution ALMA maps also revealed that the main filamentary structure is along the north-south direction
\citep{kainulainen17,hacar18}

%
%

\subsection{N$_2$H$^+$}

 
 In Figure \ref{fig:co}c, we present the intensity maps of N$_2$H$^+$ velocity-integrated 
 from 1.6 km s$^{-1}$ to 20.1 km s$^{-1}$. In other words, we integrate the emission of 
 all 7 hyperfine components.
 
The strongest emission of N$_2$H$^+$ ($J=1-0$) comes 
near the position of FIR 4 (HOPS108)
at (R.A. [J2000], Dec. [J2000])=(\timeform{5h35m26.5s}, \timeform{-5D10'00"}).
The position of the  N$_2$H$^+$ peak agrees well with those of $^{13}$CO and C$^{18}$O.
In the following, we refer to the central compact molecular clump detected by N$_2$H$^+$ as the N$_2$H$^+$ clump.  
The clump contains FIR4/5.
The FIR 3 protostar is located at the north-east edge of the N$_2$H$^+$ clump.
The VLA 13 protostar is located at the south-west edge of the clump.
This  N$_2$H$^+$ clump corresponds to core No. 10 of \citet{tatematsu08}.
The N$_2$H$^+$ clump has a couple of spines which are indicated with dashed lines in Figure \ref{fig:co}c. 
This region is also observed with ALMA in N$_2$H$^+$ \citep{hacar18}.  
The overall distribution of the N$_2$H$^+$ emission 
agrees well with the ALMA map. For example, the positions of the spines are in good agreement 
with the structures seen in the ALMA image.   The ALMA map revealed that the FIR 4 clump are connected with a couple of small
filaments which can be recognized in Figure \ref{fig:co}c as spines.
Such a structure is morphologically reminiscent of the hub-filament structure
discussed by \citet{myers09}, 
although the size of the region is smaller than the ones originally discussed by \citet{myers09}.
It is worth noting  that the $^{13}$CO high velocity component of the VLA 13 outflow 
is anti-correlated with the N$_2$H$^+$ distribution.
Similar distributions are recognized in the maps of other high density tracers
(see the next subsections).
The other moderately-strong N$_2$H$^+$ emission comes from just north of the N$_2$H$^+$ clump, which deviates from 
the C$^{18}$O peak at the upper-right corner of Figure \ref{fig:co}b.

\subsection{CCS}


In Figure \ref{fig:co}d, we show the integrated intensity map of 
CCS ($J_N=8_7-7_6$).  
The overall distribution is roughly similar to that of C$^{18}$O.
This is the first detection of the CCS emission in the OMC-2 region. 
Previous observations by \citet{tatematsu08} reported no detection of CCS ($J_N=4_3-3_2$)
presumably because of lower sensitivity.  
The strongest emission comes from the upper-right corner of the observed area
(west of HOPS68), where
the $^{13}$CO and C$^{18}$O emission is strong. 
This component is likely to belong to core No. 320 of \citet{shimajiri15}. 
We also see another local peak in the northern part just above the N$_2$H$^+$ clump (east of HOPS68).
The other strong peaks of CCS are located in the N$_2$H$^+$ clump
    toward FIR 4 (HOPS108) at (R.A. [J2000], Dec. [J2000]) = (\timeform{5h35m27s}, \timeform{-5D10'00"}), and
just in the south-east part of VLA 13 (HOPS368).


\subsection{HCO$^+$ and H$^{13}$CO$^+$}


In Figures \ref{fig:co}e and \ref{fig:co}f, we present the integrated intensity maps of HCO$^+$ and H$^{13}$CO$^+$.
The HCO$^+$ distribution is elongated in the north-south direction
and its distribution resembles those of $^{13}$CO and C$^{18}$O. 
But, the position of the local peak at the northern part significantly deviates from 
those of the $^{13}$CO and C$^{18}$O (east of HOPS68) and is consistent with the N$_2$H$^+$ peak. 
H$^{13}$CO$^+$ emission basically follows the HCO$^+$ distribution
and is  more extended than the N$_2$H$^+$ distribution.

\subsection{HN$^{13}$C and H$^{13}$CN}


In Figures \ref{fig:co}g and \ref{fig:co}h we present the integrated intensity maps of HN$^{13}$C and H$^{13}$CN.
H$^{13}$CN has three hyperfine components and we integrated all three components to make the map.
Roughly speaking, the spatial distributions of these emission lines are similar to that of H$^{13}$CO$^+$, although
they are less extended. 

The strongest H$^{13}$CN emission comes from the N$_2$H$^+$ clump (at the position of FIR 4).
HN$^{13}$C has a few relatively-strong peaks at 
(R.A. [J2000], Dec. [J2000]) = (\timeform{5h35m27s}, \timeform{-5D10'17"}), (\timeform{5h35m23s}, \timeform{-5D10'9"}) and 
(\timeform{5h35m25s}, \timeform{$-$5D8'35"}).
The strongest  peak is between FIR 4 and FIR 5.
The other local peak is located between HOPS65 and HOPS368.  
Another local peak is located near HOPS68.
In the N$_2$H$^+$ map, we see spines stretching from the central N$_2$H$^+$ clump toward the south-west direction. 
The HN$^{13}$C emission appears to follow the spines seen in N$_2$H$^+$.



\subsection{DNC, N$_2$D$^+$, and DC$_3$N}


We did not detect N$_2$D$^+$ and DC$_3$N emission at the noise level of $\sim$ 0.3 K at an velocity resolution of 0.05 km s$^{-1}$.
Below we describe only the distribution of DNC.
In Figure \ref{fig:co}i, we present the velocity integrated intensity map of DNC.
Relatively strong emission of DNC comes between FIR 5 and VLA 13
and  the emission is extended toward the west direction from VLA 13 (HOPS368).



\section{Derivation of Physical Quantities}
\label{sec:derivation}

\subsection{N$_2$H$^+$ and CCS abundances}

\begin{figure}
\centering
\includegraphics[width=0.5 \textwidth, bb=-100 0 317 255]{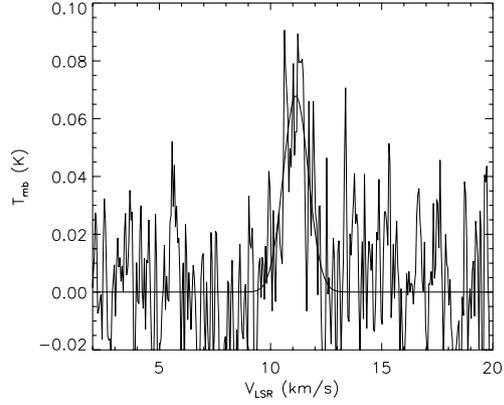}
\caption{The averaged  CCS line profile of the area which contains FIR 3/4/5 and VLA 13.
We averaged the CCS integrated intensity in the N$_2$H$^+$ clump with $\sim$ 0.17 K km s$^{-1}$.
The thick solid line is the result of the Gaussian fitting, from which
we obtained the peak intensity of 0.068 K, central velocity of 11.14 km s$^{-1}$, and line width of 1.3 km s$^{-1}$. 
}
\label{fig:ccsprofile}
\end{figure}

Here, we estimate the abundances of CCS  and N$_2$H$^+$ at several positions.  
Since the CCS emission is weak, we derive only the values  averaged in the small area 
which contains all four sources FIR 3/4/5 and VLA 13, which is indicated in dashed lines
in Figure \ref{fig:co}d.
The averaged profile is shown in Figure \ref{fig:ccsprofile}.
The derived physical quantities are summarized in Table \ref{tab:ccsn2hp}.

We compute CCS column densities assuming that the emission is optically thin. 
The excitation temperature is adopted to be 5 K.
The CCS fractional abundance does not vary significantly from position to position, and
stays at about $10^{-12}$.
Here, we derived the column density of H$_2$ from the {\it Herschel} H$_2$ map.
The rms noise level of the CCS data was about 0.036 K km s$^{-1}$ for the angular resolution of 32$"$.5, corresponding to 
$\sim 0.5 \times 10^{11}$ cm$^{-2}$. Therefore, the CCS integrated intensities at the positions of FIR3/4/5 and VLA13 were from 4 to 7 $\sigma$.
According to the chemical evolution calculations by \citet{suzuki92} and \citet{marka12},
such a small abundance can be achieved at the times of $10^2$ years and $10^5$ years.
According to \citet{marka12}, the time evolution of the CCS abundance is not so different
for the temperatures from 10 K to 25 K, whereas for much higher gas temperatures,  
the chemical time scale may be shorter due to faster chemical reaction \citep[see e.g.,][]{shimoikura18}.  
In the FIR 4 region, the gas temperature is likely to be $\sim$ 20 K. Thus, the chemical timescales
should be comparable to those in dark clouds with $\sim$ 10 K.
Since the protostars already formed in this region, the evolution time of this region is 
likely to be  $\sim 10^5$ years, which is comparable to the typical lifetime of the Class I protostars.

\begin{figure}
\centering
\includegraphics[width=0.18 \textwidth, bb=-200 0 400 312]{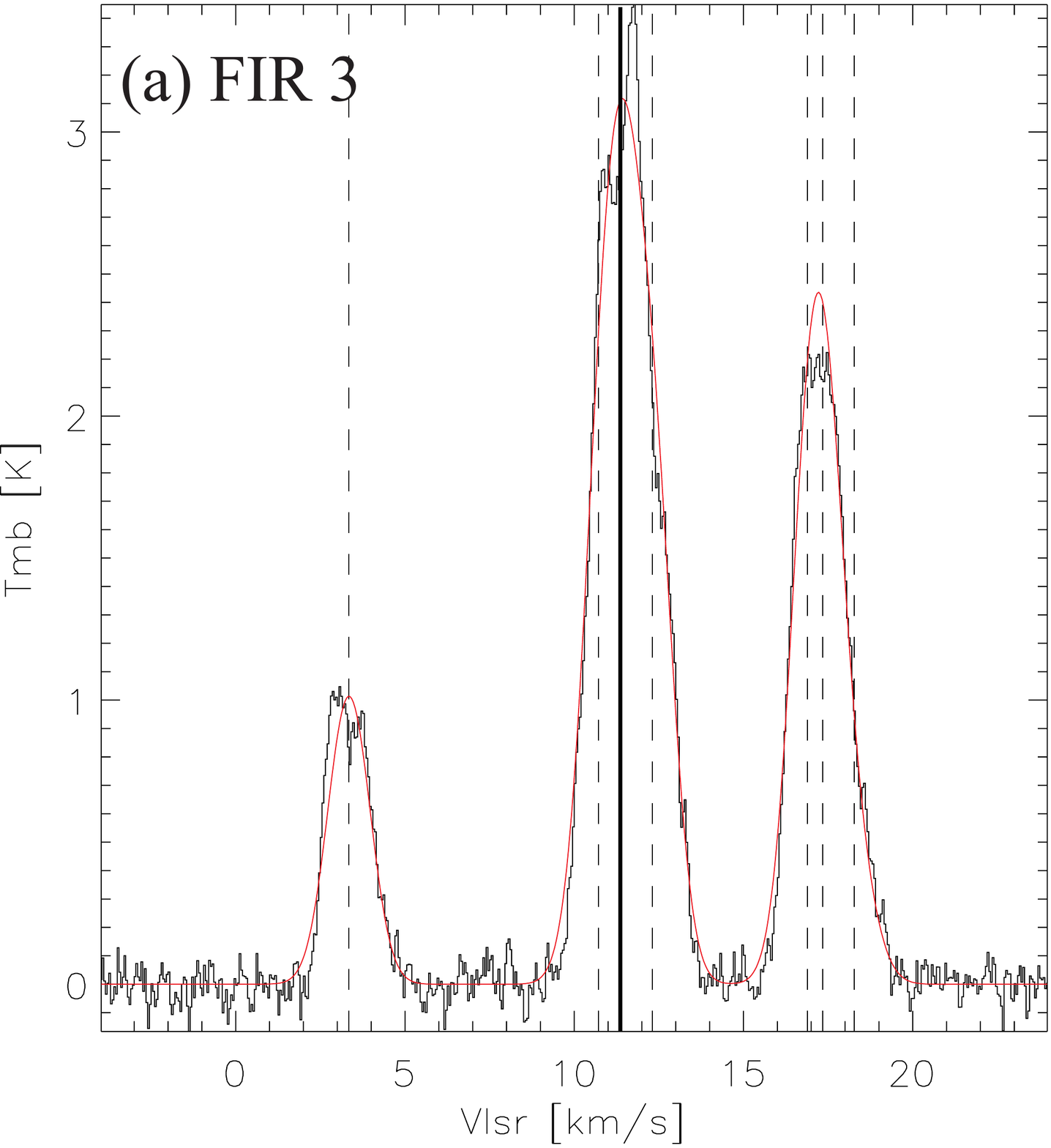}
\includegraphics[width=0.18 \textwidth, bb=-200 0 400 312]{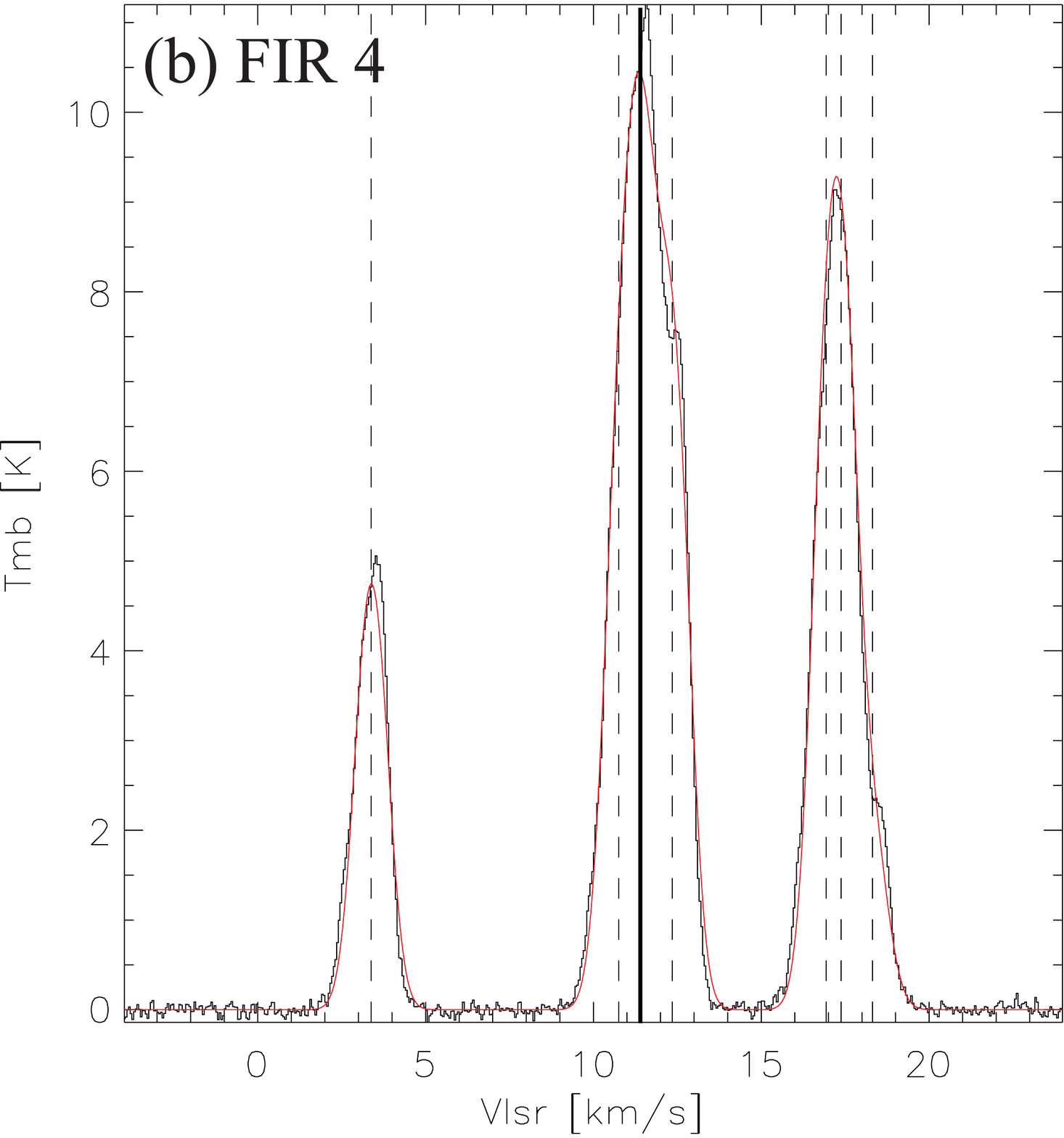}
\includegraphics[width=0.18 \textwidth, bb=-200 0 400 312]{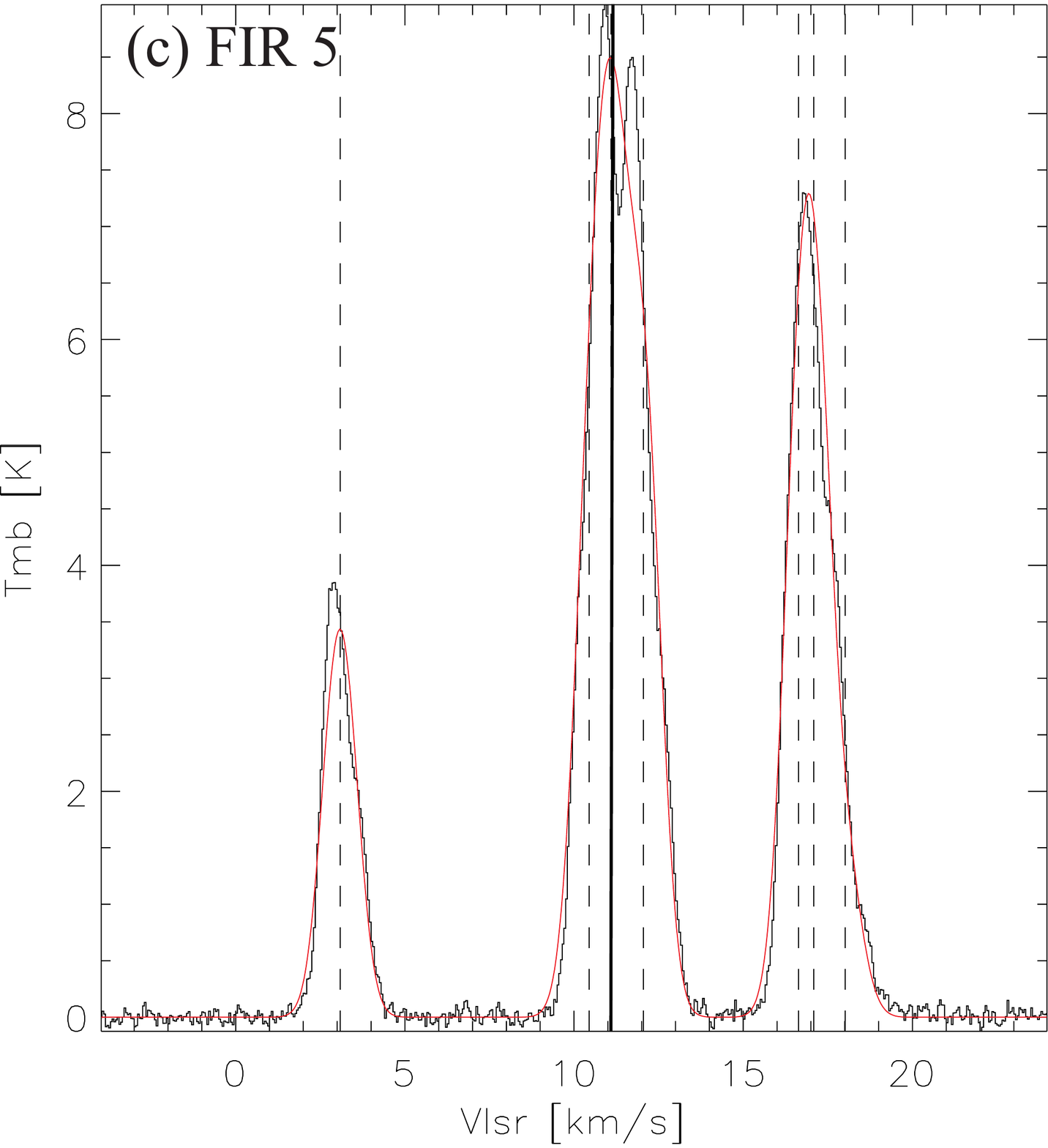}
\includegraphics[width=0.18 \textwidth, bb=-200 0 400 312]{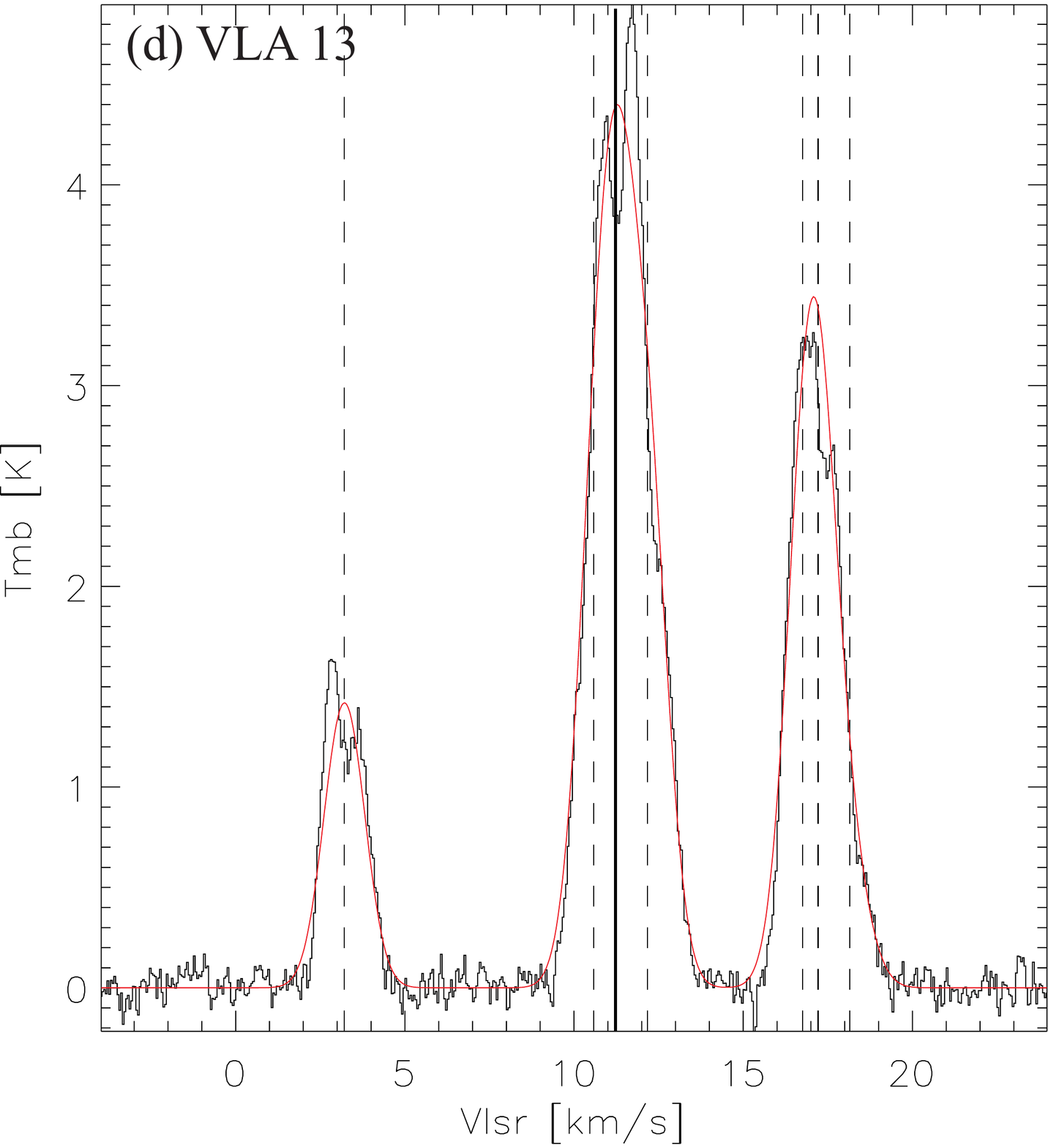}
\caption{Hyperfine fit of N$_2$H$^+$ line profiles at (a) FIR 3, (b) FIR 4, (c) FIR 5, and (d) VLA 13.
The red curves show the results of the fitting.  The frequencies of the seven hyperfine
components are indicated by lines. The main component is indicated by a solid line in each panel.
}
\label{fig:hyperfinefit}
\end{figure}

For N$_2$H$^+$, we apply the hyperfine fit to derive the physical parameters.
The results of the hyperfine fits at FIR 3, FIR 4. FIR 5, and VLA 13 are shown in Figure \ref{fig:hyperfinefit}.
We used the same fitting program as that of \citet{tanaka13}, in which 
the excitation temperatures and velocity widths are assumed to be identical for the seven hyperfine components.
For simplicity, we assume that only one component exists along the line of sight, although
this assumption seems not to be appropriate at least for FIR 3 and VLA 13, both of which have at least two
components. 
The fractional abundance  of N$_2$H$^+$ is relatively high in this region.
It is highest at FIR 4.
The excitation temperature of  N$_2$H$^+$ is relatively high at $\sim 20 $ K.
It is strongest at the position of VLA 13.
In the starless cores in the southern part of Orion A, the N$_2$H$^+$ excitation temperatures are as low as $5\sim$10 K \citep{tatematsu14}. 
In Oph B2 region where only 4 protostars are identified, the N$_2$H$^+$ excitation temperatures stay as low as 5 K \citep{friesen10}.
Other examples of the low excitation temperatures in starless regions can be found in \citet{james04} and \citet{tanaka13} toward Oph A and Serpens South, respectively.
These previous studies indicate that in  the regions where star formation is not active, the excitation temperatures of N$_2$H$^+$ stay as low as 5 $-$10 K.
Thus, the high excitation temperature in the FIR 4 region may indicate the existence of protostars in this region.
The line widths are about 1 km s$^{-1}$ for all the positions, which is comparable to that of CCS.

\begin{table}
  \tbl{Physical Parameters of CCS and N$_2$H$^+$ toward FIR 3, FIR 4, FIR 5, and VLA 13}{%
  \begin{tabular}{llllll}
  \hline
  & Area & FIR 3 & FIR 4 & FIR 5 & VLA 13 \\ \hline
${\rm N_{H2}(\times 10^{23} \, cm^{-2})}$ & 2.3 & 1.5 & 2.8 & 2.5 & 0.69 \\ \hline
${\rm N_{CCS}(\times 10^{11} \, cm^{-2})}$ & 2.8 & 2.2 & 3.4 & 3.1 & 2.6 \\
${\rm f_{CCS}(\times 10^{-12})}$    & 2.0 & 1.5 & 5.0 & 1.2 & 3.8 \\  
$\Delta V$  (km s$^{-1}$)    & 1.3 &  &  & &  \\      
${\rm \tau_{CCS}}$      & 0.12  &  &  &  &  \\ \hline
${\rm N_{N_2H^+}(\times 10^{14} \, cm^{-2})}$ & & 5.8 & 2.9 & 2.3 & 1.2 \\
${\rm T_{ex, N_2H^+} (K)}$ & & 16.0 $\pm$ 0.9 & 21.8 $\pm$ 0.1 & 23.1 $\pm$ 0.3& 26.2 $\pm$ 1.5 \\
$\Delta V$  (km s$^{-1}$) & & 1.4 $\pm$ 0.1& 1.1 $\pm$ 0.1& 1.1 $\pm$ 0.1& 1.4 $\pm$ 0.1 \\
${\rm f_{N_2H^+}(\times 10^{-10})}$ & & 3.8 & 10.3 & 8.9 & 1.7 \\
${\rm \tau_{N_2H^+}}$ & & 0.8 $\pm$ 0.1 & 2.9 $\pm$ 0.1 & 1.9 $\pm$ 0.1 & 0.6 $\pm$ 0.1 \\ \hline\hline
  \end{tabular}}\label{tab:ccsn2hp}
\end{table}

\subsection{The HN$^{13}$C/H$^{13}$CN Ratio}

Previous observations revealed that the  [HNC]/[HCN] ratio varies from region to region.
For example, in  warm dense gas in OMC-1, it is measured to be much smaller than unity \citep{goldsmith81,goldsmith86}.
On the other hand, in cold dense gas, \citet{tennekes06} derived a somewhat large 
ratio of 3 $-$ 4 toward Cha-MM1.
\citet{colzi18} showed that for high-mass starless cores, the [HNC]/[HCN] ratio stays at around unity, whereas it decreases
for more-evolved objects such as high-mass protostellar cores ($\sim$ 0.5). 
\citet{jin15} found that the [HNC]/[NCN] ratio increases from IRDCs to UC HIIs.
These studies indicate that  the [HNC]/[HCN] ratio depends on the physical conditions.
The  [HNC]/[HCN] ratio may be a useful indicator to guess the physical states of star-forming regions.
However, it still remains uncertain which mechanism is more responsible for determining the [HNC]/[HCN] ratio.
Here, we examine the spatial distribution of the [HNC]/[HCN] ratio in the FIR 4 region to gain additional information to understand the HNC/HCN chemistry.

\citet{hirota98} proposed that the [HNC]/[HCN] ratio depends on the gas temperature and rapidly decreases as the temperature exceeds the critical value of 24 K, 
above which  a neutral-neutral reaction of HNC + H $\rightarrow$ HCN + H is responsible for destroying HNC \citep[see also][]{schilke92,talbi96}.
\citet{colzi18} discussed that the small ratio for high-mass protostellar cores may be due to the effect of selective destruction of HNC by the reactions of HNC + H $\rightarrow$ HCN + H
and HNC + O $\rightarrow$ NH + CO, suggesting the existence of the threshold temperature for the selective destruction of HNC, although
future laboratory experiments are needed to accurately estimate the reaction rate with O at low temperatures to understand the temperature dependence of the [HNC]/[HCN] ratio.
\citet{loison14} pointed out that in the presence of carbon atom in gas phase,
C + HNC reaction also prevents the [HNC]/[HCN] ratio from reaching unity.  
On the other hand, \citet{aguado17}  discussed the effect of FUV radiation which can selectively photodissociates HNC, 
because the photodissociation cross section is larger for HNC, leading to  smaller [HNC]/[HCN] ratios for stronger FUV.
However, the parameter space is not fully explored for the effects of the FUV radiation. Further investigations would be needed.

\begin{figure}
\centering
\includegraphics[width=0.4 \textwidth,bb=0 -100 452 700]{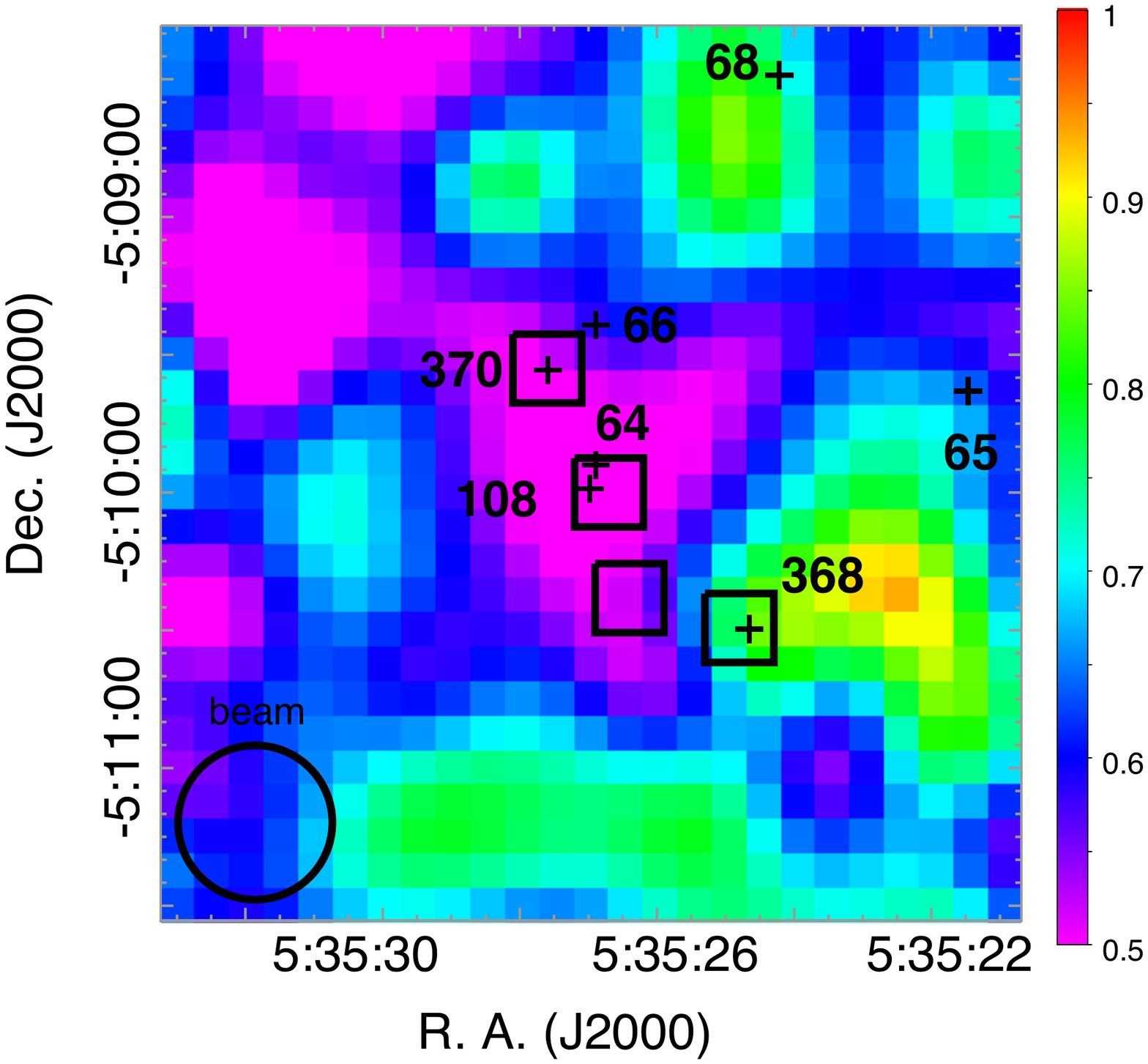}
\caption{[HN$^{13}$C]/[H$^{13}$CN] ratio toward the FIR 4 region. In the present paper, we consider that
the [HN$^{13}$C]/[H$^{13}$CN] ratio is identical to the [HNC]/[HCN] ratio.
}
\label{fig:hnc-to-hcn}
\end{figure}

In Figure \ref{fig:hnc-to-hcn} we show the  spatial distribution of the [HNC]/[HCN] ratio of the FIR 4 region.
Here, we assume that [HNC]/[HCN] ratio is identical to [HN$^{13}$C]/[H$^{13}$CN] ratio
and we derived the column densities of HN$^{13}$C and H$^{13}$CN by assuming that both emission
is optically thin.
Here, the excitation temperature is set to 20 K to compute the column densities.  
We note that  the column densities are insensitive to the assumed excitation temperature and
they varies only within errors of 30 \% in the range of 10 K to 30 K.
In the N$_2$H$^+$ clump, the [HNC]/[HCN] ratio is estimated to be  $\sim$ 0.5.
It is smaller than the value at Cha-MMS 1 which has the ratio of 3$-$4, but
not extremely small such as in the OMC-1 hot core. 
It is somewhat higher in the envelope of the N$_2$H$^+$ clump.
The average ratio in the clump seems to be consistent with the values ($\sim 0.5$) in the high-mass protostellar objects derived by \citet{colzi18}.

As we mentioned above, there are at least two effects discussed for reducing the [HNC]/[HCN] ratio \citep{hirota98,colzi18,aguado17}. 
One  is that the gas temperature exceeds the critical temperature for the HNC formation.
In this case, the achieved [HNC]/[HCN] ratio tends to be very small ($\lesssim 0.1$) for higher temperature ($T \gg 24 $K).
The average dust temperature of the N$_2$H$^+$ clump is estimated to be about 26 K \citep{lombardi14}, which is 
somewhat larger than or at least comparable to the critical temperature of 24 K.  
Second is the FUV radiation which can selectively photodissociate HNC, resulting in the smaller [HNC]/[HCN] ratio.
Recently, Ishii et al. (2018) estimated the dimensionless FUV strength $G_0$  from the {\it Herschel} 70 $\mu$m  data and
they found that the $G_0$ parameter of the FIR 4 region exceeds 100 and peaked at around FIR 3 ($\sim$ 7000) and VLA 13 ($\sim 1400$).
If the UV radiation is more responsible for determining the spatial distribution of the ratio, it should be lower at around VLA 13, similarly to at FIR 3.  
In contrast, our derived ratio is somewhat larger at VLA 13 than in the N$_2$H$^+$ clump and FIR 3, and thus seems to be inconsistent with the effect of the FUV radiation.
Although further understanding of the HNC and HCN chemistry would be needed, we tentatively conclude that the effect of the temperatures comparable to the critical value ($T > 24$ K)
is more responsible for the observed [HNC]/[HCN] ratio.

If the [HNC]/[HCN] ratio is a good indicator of the temperature in the dense regions, 
our [HNC]/[HCN] image might provide an evidence of the  outflow-dense gas interaction in this region.
When the FIR 3 outflow interacts with the dense molecular gas in the FIR 4 region, 
the dynamical compression heats up the gas. As a result, 
the [HNC]/[HCN] ratio in the FIR 4 clump may be somewhat reduced. At the positions of the 
outflow lobes of the FIR 3 and VLA 13 outflows (see the distribution of the outflow lobes in HCO$^+$ 
(figure \ref{fig:hcooutflow} in the next section) and
$^{12}$CO \citep{takahashi08}),  the [HNC]/[HCN] ratio is smaller than the values in the surrounding parts. 
These features of the [HNC]/[HCN] ratio may be a sign of the outflow-dense gas interaction.
In addition, the obtained [HNC]/[HCN] ratio in the FIR 4 clump might be inconsistent with 
the scenario that FIR 4 is a hot core \citep{ceccarelli10,kama13}, 
since the [HNC]/[HCN] ratio in the OMC-1 hot core is much smaller 
than the value in  FIR 4 \citep{goldsmith81,goldsmith86}.

\section{Molecular Outflows}
\label{sec:outflow}

In the observed area, 7 protostars are previously identified, which are listed in Table \ref{tab:protostars}.
5 of them are classified as Class 0/I. Such  protostars often drive powerful molecular outflows.
\citet{takahashi08} performed the outflow survey in CO ($J=3-2$) toward Orion A and in this region,
they detected three outflows [FIR 2 (HOPS68), FIR 3 (HOPS370), VLA 13 (HOPS368)].
Here we identify the outflows driven on the basis of the $^{13}$CO and
HCO$^+$ images.

\begin{figure}
\begin{center}
\includegraphics[width=0.2 \textwidth,bb=-100 0 432 500]{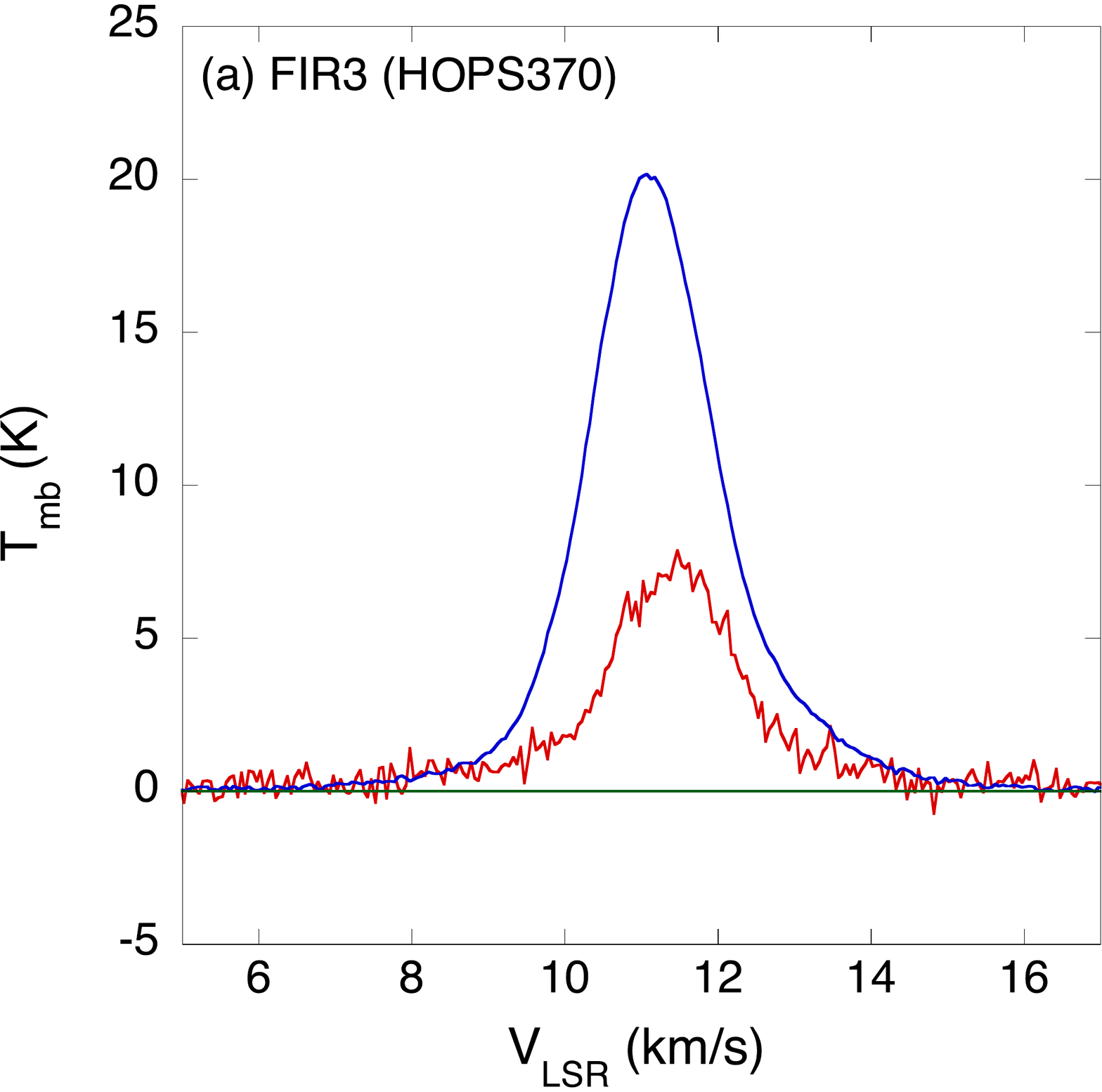} \hspace{0.5cm}
\includegraphics[width=0.2 \textwidth,bb=-100 0 448 500]{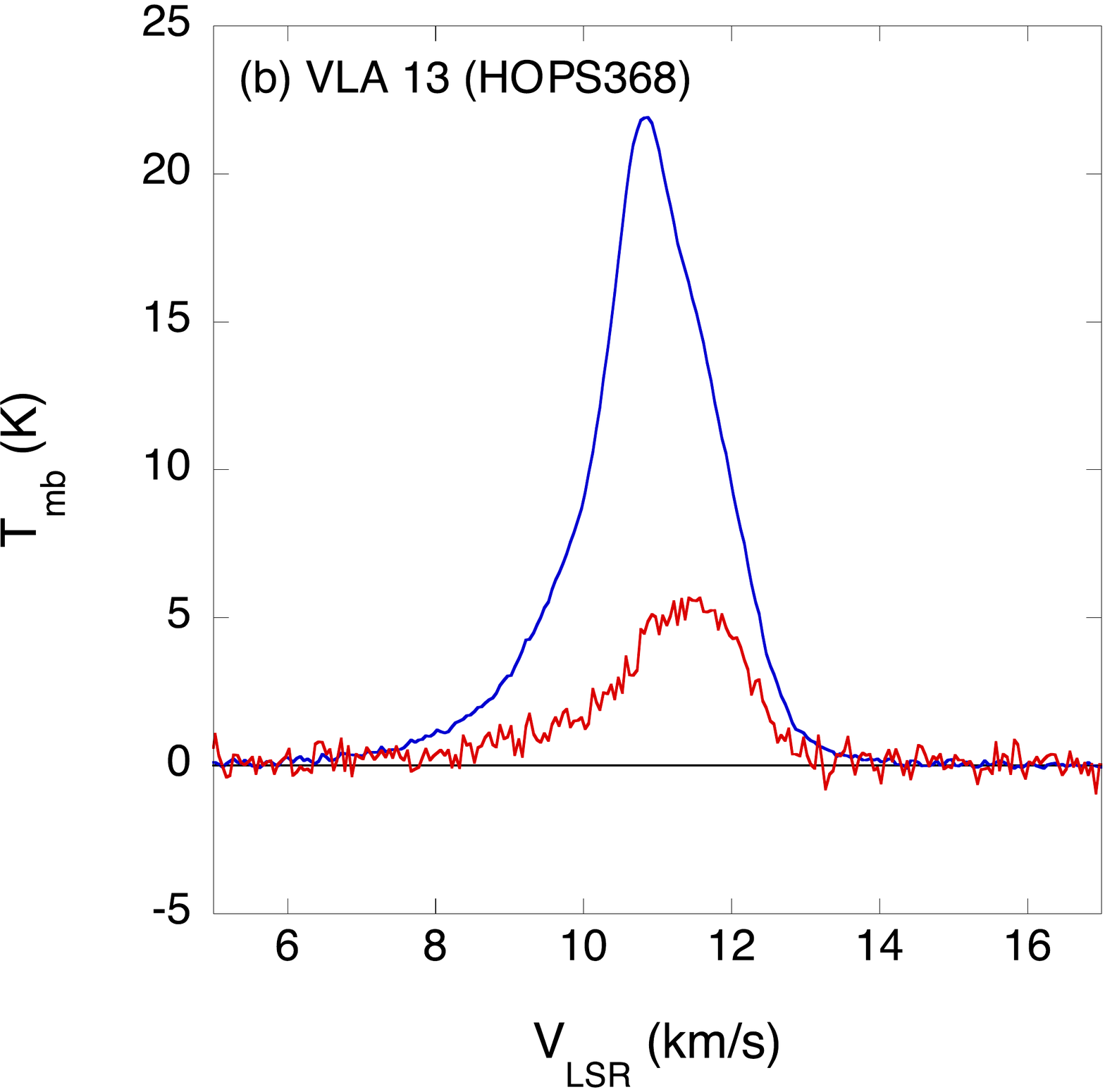}
\includegraphics[width=0.2 \textwidth,bb=-100 0 448 500]{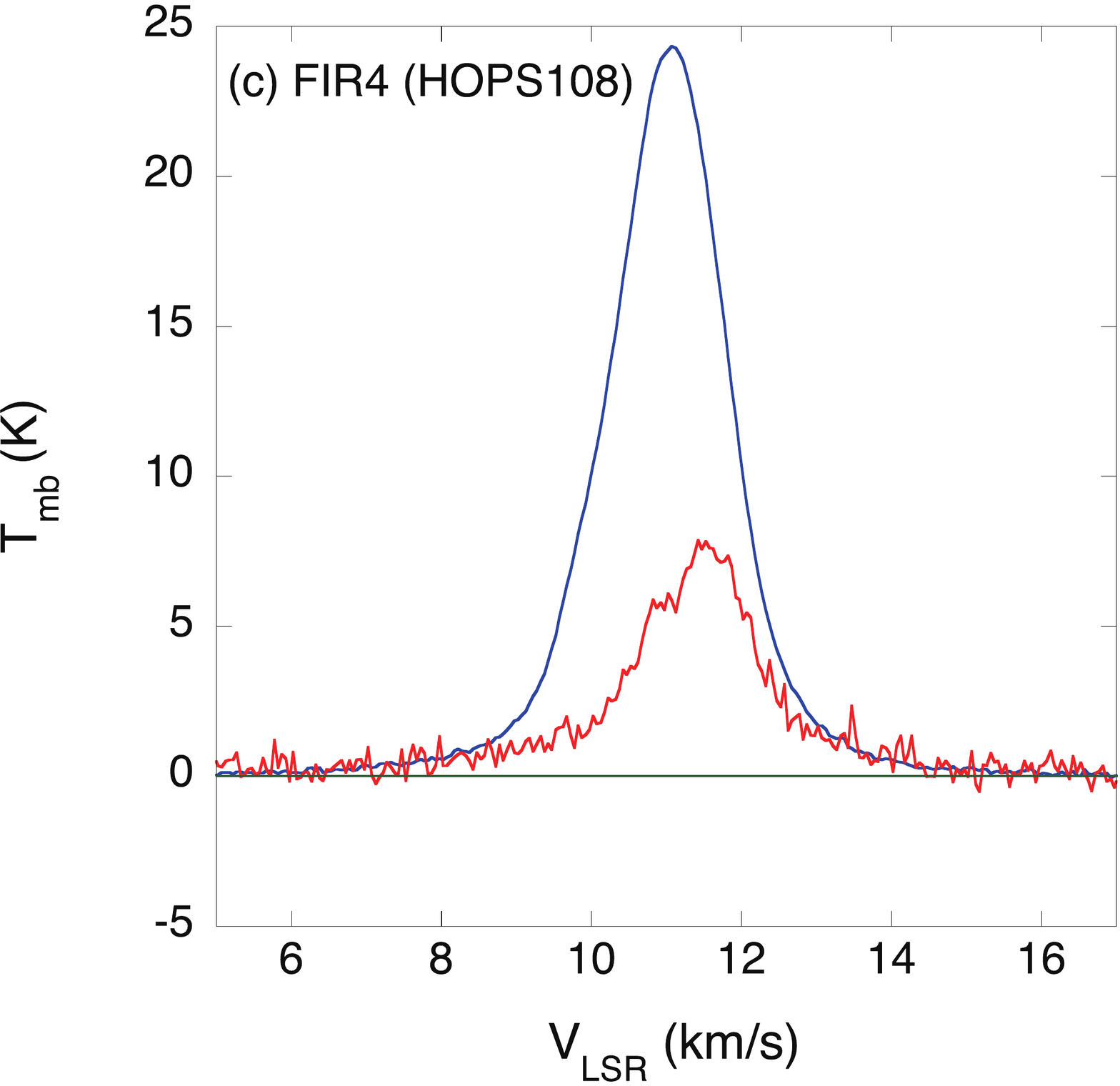}
\end{center}
\caption{The $^{13}$CO and HCO$^+$ profiles towards (a) FIR 3, (b) VLA 13,  and (c) FIR 4.
The blue and red curves indicate the $^{13}$CO ($J=1-0$) and HCO$^+$ ($J=1-0$) line profiles, respectively.
We note that the effective angular resolution of the HCO$^+$ profiles is 24.$"$9, the original resolution.}
\label{fig:profile}
\end{figure}

\begin{figure}
\begin{center}
\includegraphics[width=0.2 \textwidth,bb=0 0 476 422]{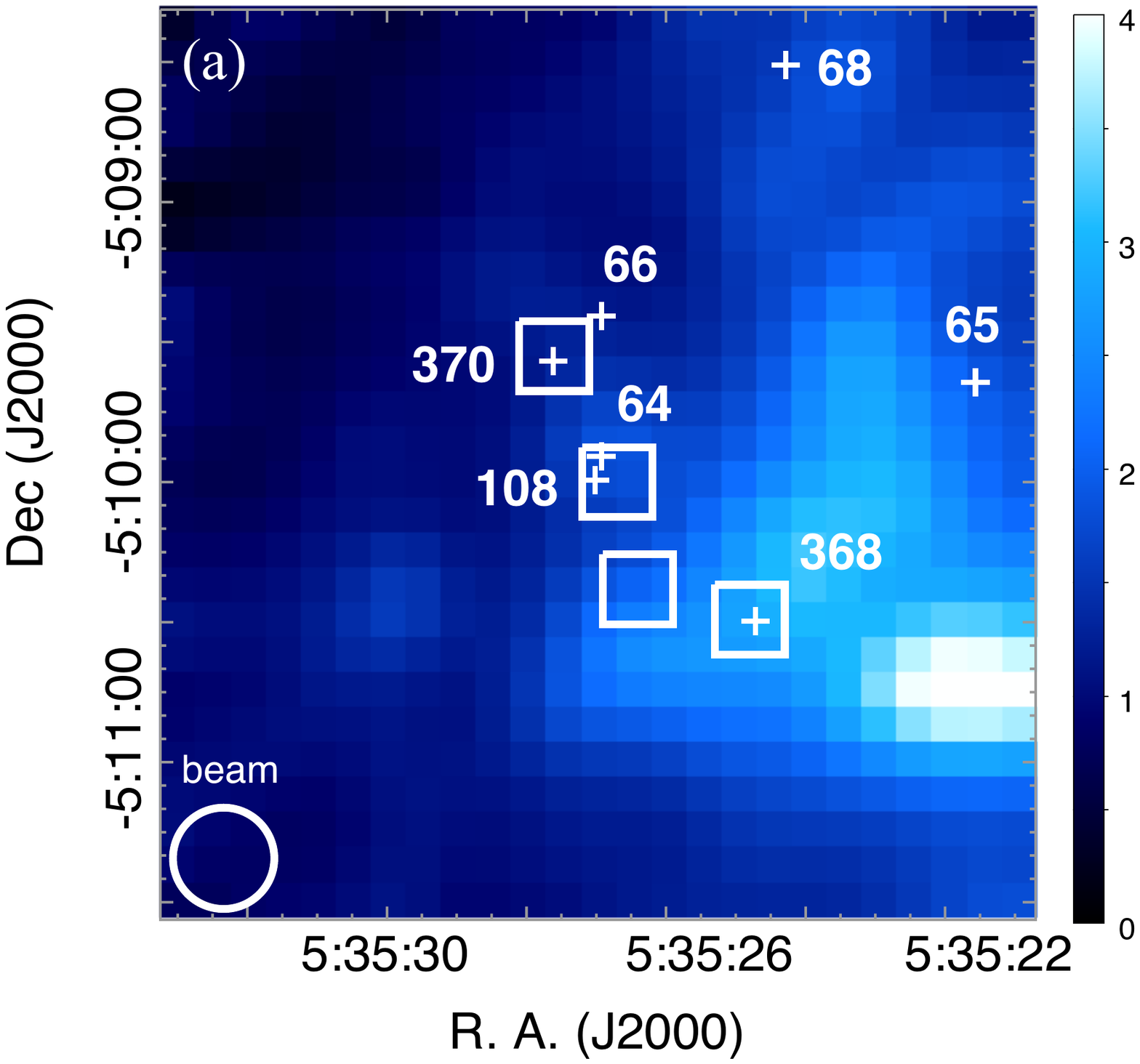}
\includegraphics[width=0.2 \textwidth, bb=0 0 475 421]{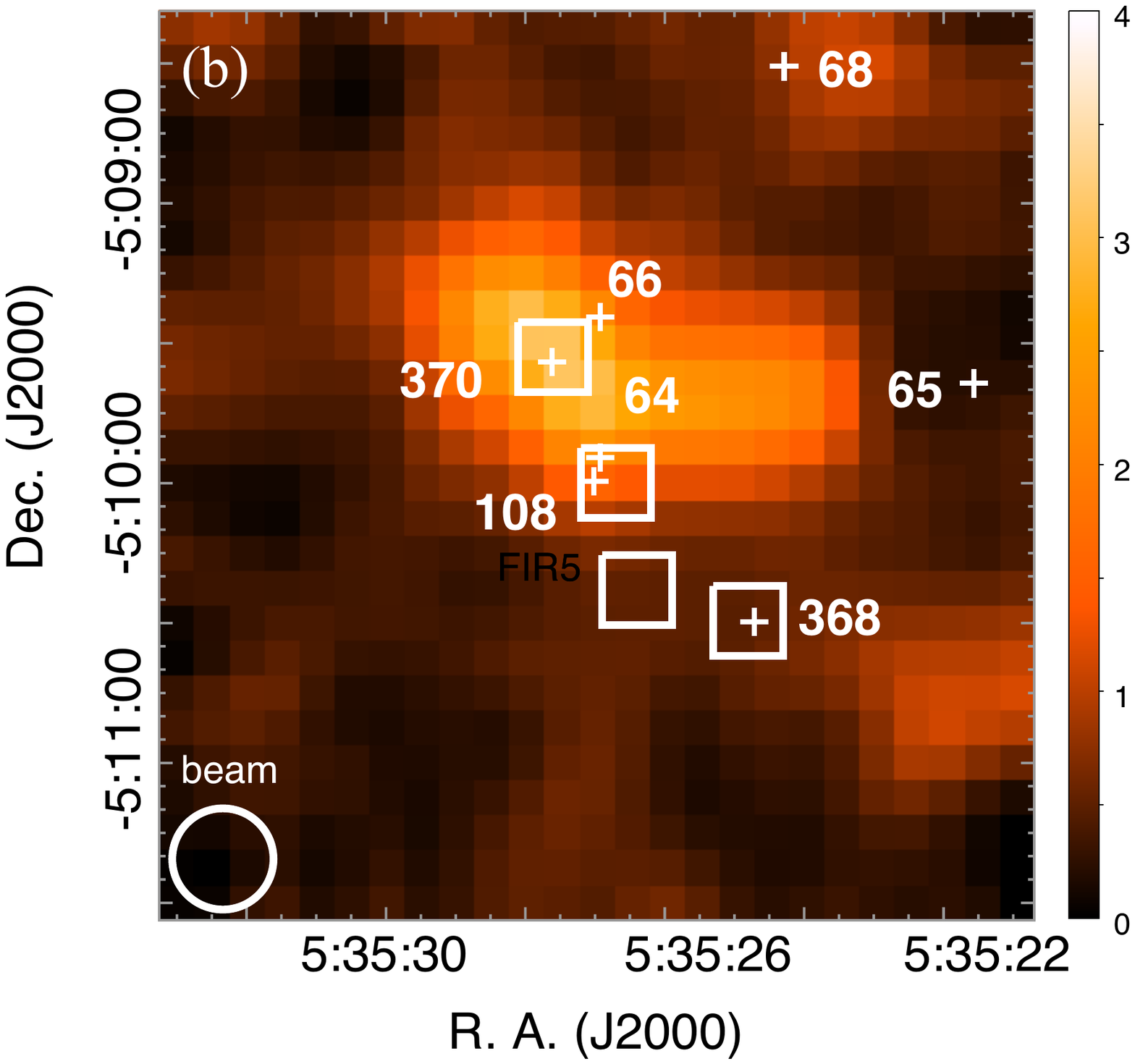}
\includegraphics[width=0.2 \textwidth, bb=0 0 472 419]{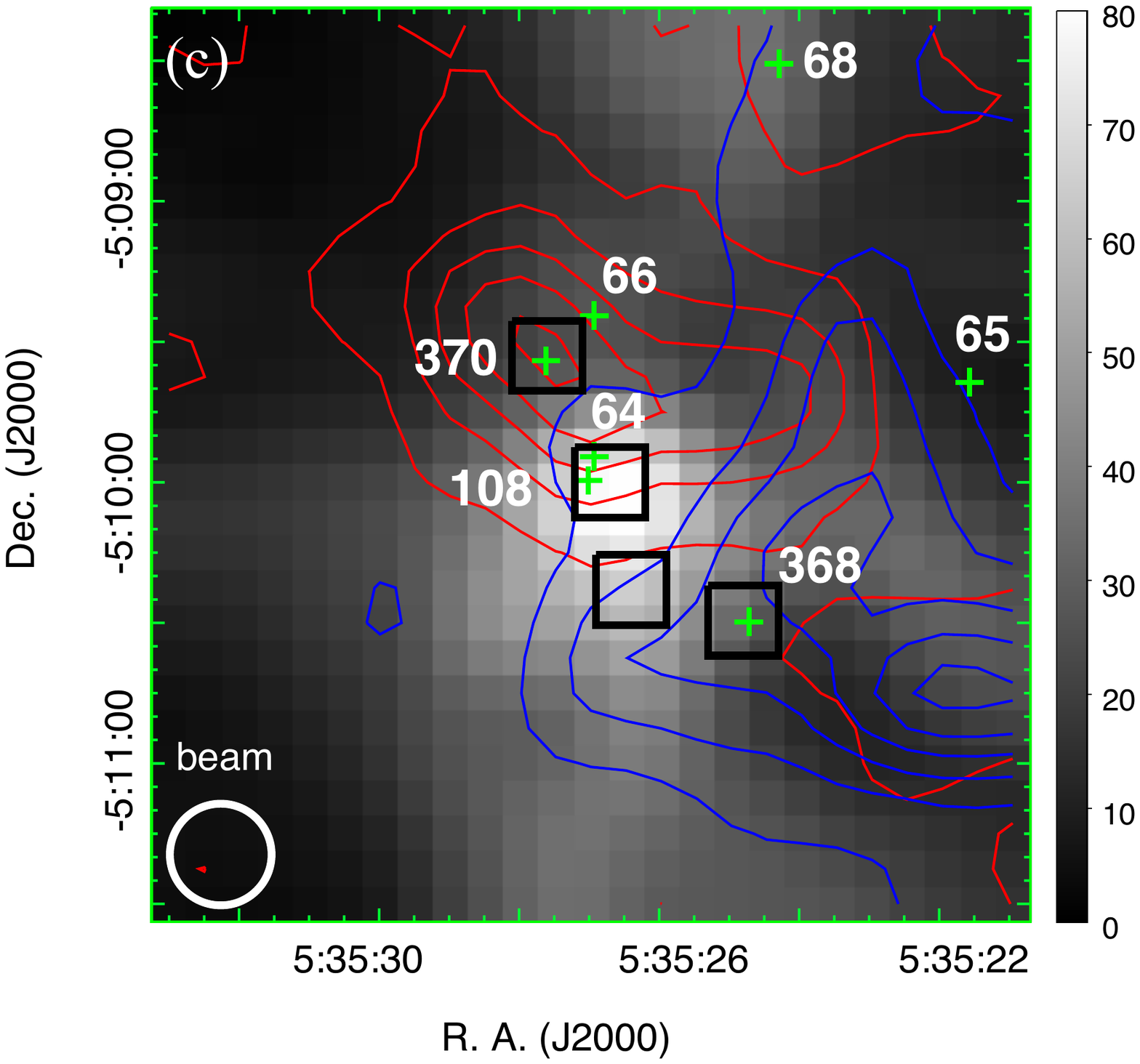}
\end{center}
\caption{Distributions of high velocity components of $^{13}$CO ($J=1-0$)  (a) blueshifted and 
(b) redshifted components in the FIR 3/4/5 region. The blueshifted and redshifted
components are integrated over 6 $-$ 9 km s$^{-1}$ and 13 $-$ 15 km s$^{-1}$, respectively.
(c) the $^{13}$CO high velocity components overlaid on the N$_2$H$^+$ integrated intensity image.
The positions of the protostars listed in Table \ref{tab:protostars} are indicated with green crosses. 
The red and blue contours start at 1.5 K km s$^{-1}$ and 0.6 K km s$^{-1}$ with an interval of 0.5 K km s$^{-1}$ 
and 0.6 K km s$^{-1}$, respectively.}
\label{fig:cooutflow}
\end{figure}

\begin{figure}
\begin{center}
\includegraphics[width=0.25 \textwidth, bb=0 0 514 443]{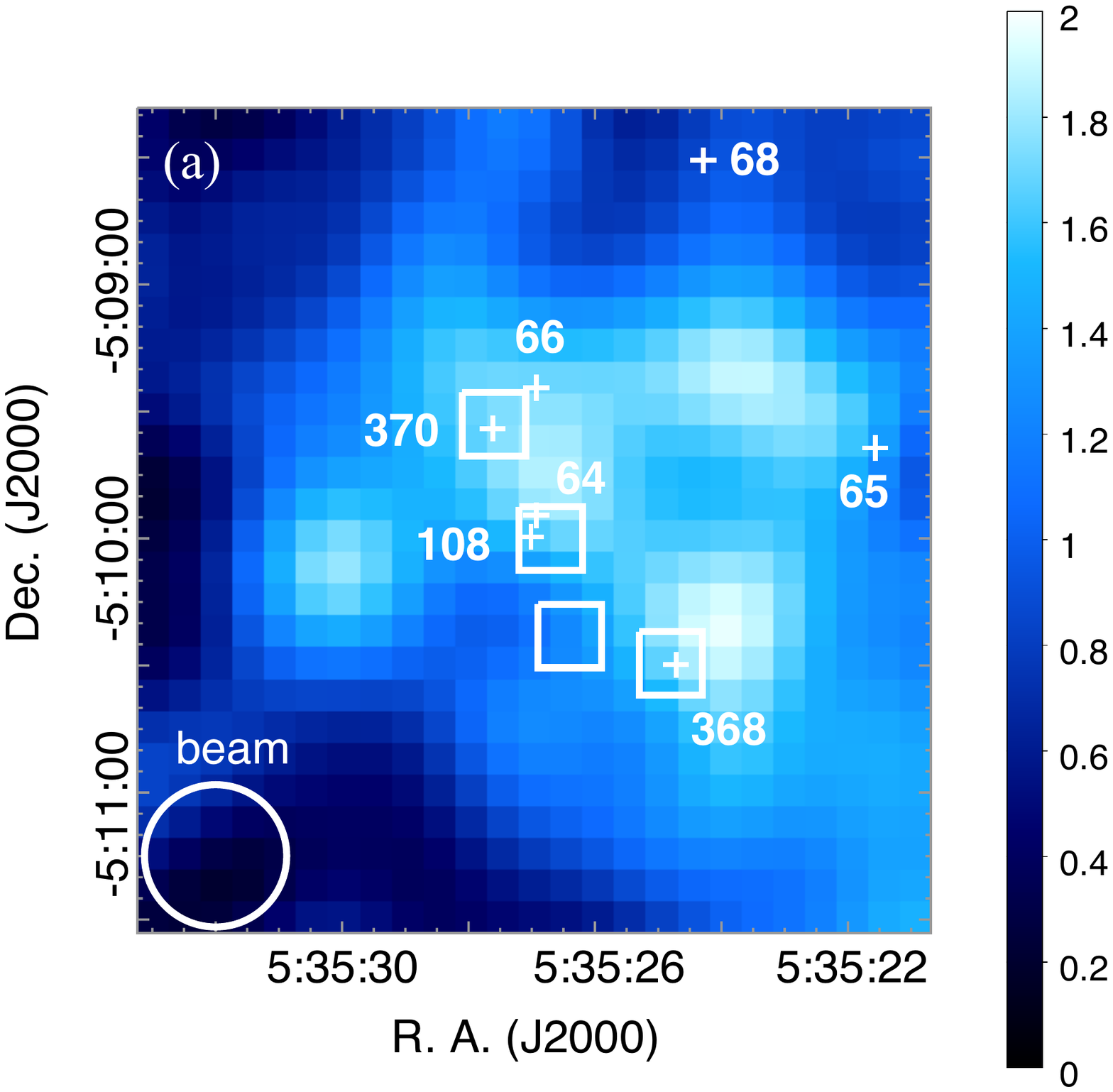}
\includegraphics[width=0.25 \textwidth, bb=0 0 516 443]{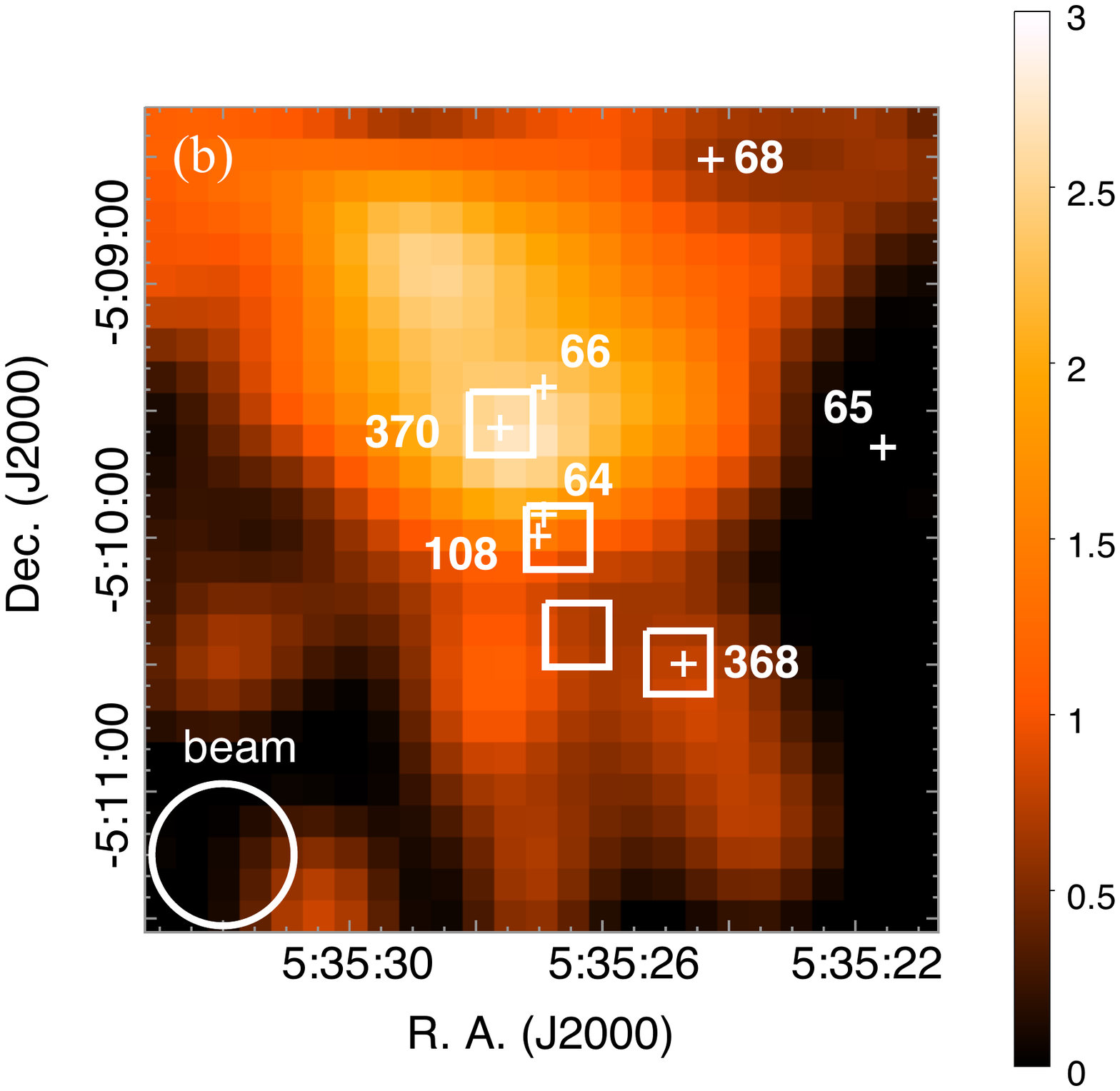}
\includegraphics[width=0.28 \textwidth, bb=0 -70 528 633]{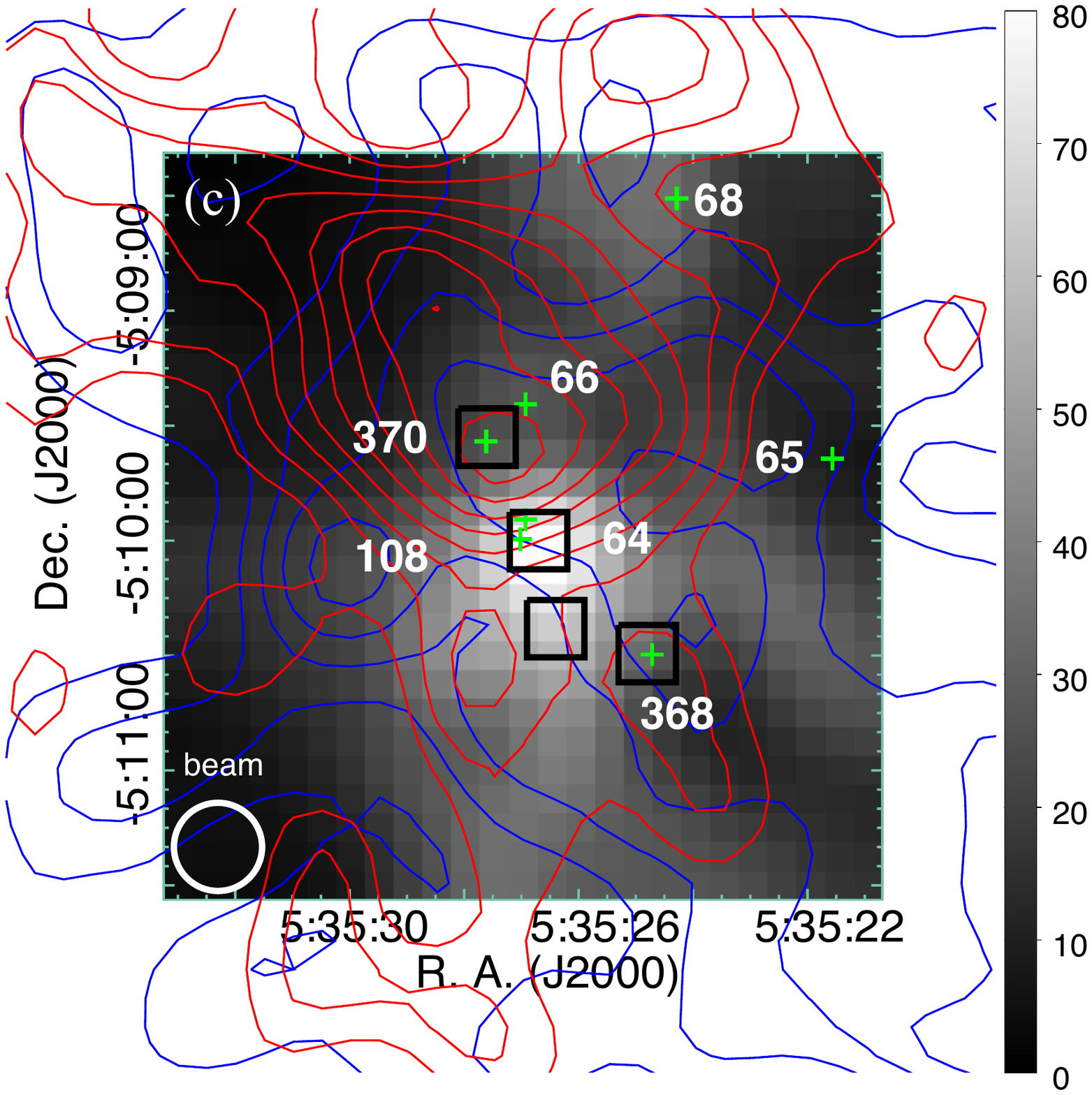}
\end{center}
\caption{Same as Figure \ref{fig:cooutflow} but for HCO$^+$. 
The velocity range for the integration is the same as those of $^{13}$CO.
The maps taken with T70 is slightly wider than those of FOREST.  Therefore, in panel (c), 
the belushifted and redshifted emission is distributed outside the N$_2$H$^+$ map.
We smoothed the HCO$^+$ data to improve the signal-to-noise ratios of the blueshifted and redshifted components.
The resultant effective angular resolution was about 34$"$.
Both red and blue contours start at  0.4 K km s$^{-1}$  with an interval of 0.3 K km s$^{-1}$.
The images are smoothed to improve the signal-to-noise ratios, and the effective resolutions are
$\sim 30"$.
}
\label{fig:hcooutflow}
\end{figure}

In Figures \ref{fig:profile}, we show the line profiles of $^{13}$CO and HCO$^+$  toward the three representative positions, 
FIR3, VLA13, and FIR 4.
Both the $^{13}$CO and HCO$^+$ line profiles at the three position have high-velocity wings which may originate from
molecular outflows.  
In Figures \ref{fig:cooutflow}a and \ref{fig:cooutflow}b, 
we show the blueshifted and redshifted components of the $^{13}$CO emission, respectively. 
We also show the blueshifted and redshifted $^{13}$CO emission overlaid on the N$_2$H$^+$ 
integrated intensity image in Figure \ref{fig:cooutflow}c. 
For comparison, 
in Figure \ref{fig:hcooutflow}a, \ref{fig:hcooutflow}b, \ref{fig:hcooutflow}c, we present the same figures as 
Figures \ref{fig:cooutflow}a, \ref{fig:cooutflow}b, and \ref{fig:cooutflow}c, respectively, but for HCO$^+$.
The HCO$^+$ emission is another good tracer of outflows in OMC-2/3 \citep{aso00}.

We adopt the systemic velocity of 11.3 km s$^{-1}$ which is measured in H$^{13}$CO$^+$ ($J=1-0$) \citep{shimajiri08}.
This velocity is in good agreement with the centroid velocity determined from CCS, 11.14 km s$^{-1}$.
First, previous observations suggest that the FIR 3 (HOPS370) protostar is a outflow source whose
axis (NW-SE direction) is roughly on the plane of sky \citep{takahashi08,shimajiri08}. Therefore, 
both the redshifted and blueshifted components are seen in $^{12}$CO ($J=1-0$)
[see Figure 1 of \cite{shimajiri08}].
In our $^{13}$CO maps, the redshifted components (NW-SE direction) are more prominent 
and the blueshifted component is very vague.
The redshifted components are also clearly seen in HCO$^+$.
The southern tip of the redshifted component appears to reach FIR 4 for both the $^{13}$CO and HCO$^+$ maps, 
indicating the outflow-dense gas interaction.
We note that the velocity range for the integration in $^{13}$CO is different from that of \citet{shimajiri08}. 

In the $^{13}$CO map, the redshifted component extended from FIR 3 in the east-west direction is prominent.
This component can vaguely be recognized in the HCO$^+$ map, appearing that this component is driving from HOPS66.
On the other hand, the blueshifted component is not seen. It might be overlapped by the FIR 3 component.

In the upper-right corner of the figures, we see blueshifted and redshifted components from the FIR 2 (HOPS 68) outflow.
This outflow is already reported by  \citet{takahashi08} who suggested that the outflow axis is along the north-south direction. 
In the $^{13}$CO maps, both redshifted and blueshifted components  are seen in the southern part, and the redshifted component is somewhat stronger.
In the HCO$^+$ map, we see relatively strong redshifted component in the northern part of FIR 2, which is outside the mapping area of the FOREST observations
(see Figure \ref{fig:hcooutflow}).
On the other hand, we do not find clear blueshifted lobe in HCO$^+$.
Since both components in $^{13}$CO are seen in the southern part, the outflow axis may be nearly on the plane-of-sky.

Near FIR 4 and FIR 5, the strong redshifted component which is extended to south from FIR 4 can be seen in the HCO$^+$ image. 
This component might be the outflow from FIR 4 (HOPS 64).  In fact, the compact blueshifted counterpart is distributed just north of this protostar,
although this blueshifted component might be the outflow from FIR 3.
\citet{takahashi08} also detected broad wings toward FIR 4 in CO ($J=3-2$). 
Our detected components in $^{13}$CO and HCO$^+$ may be related to their detected wings.

Finally, we see high velocity components near VLA 13 (HOPS368).
A faint redshifted component is seen in the HCO$^+$ emission in the southwest direction of VLA 13.
On the other hand, the blueshifted component is seen just in the north of VLA 13.
The outflow axis appears to be somewhat tilted in the NE-SW direction in the plane-of-sky.
In the $^{13}$CO maps, we see both redshifted and blueshifted components in the western part.
We note that the axis of the outflow seen in $^{12}$CO ($J=1-0$) also appears to be slightly tiled in the NE-SW direction
(see Tanabe et al. 2018), consistent with our interpretation. 
Since the dense gas is not distributed in the NE-SW direction, there is a possibility 
that the VLA 13 outflow might blow away the dense gas in this area, and dynamically  compressed the dense clump of FIR 4.
The outflow identification in Orion A using $^{12}$CO ($J=1-0$) is discussed in more details by Tanabe et al. (2018).
It is worth noting that the dense gas distribution is anti-correlated with the distribution of the outflow component.
However, this anti-correlation is not strong evidence of the outflow interaction.

In total, we confirmed 4 known outflow candidates  including marginal detection by \citet{takahashi08}  and 
1 new possible outflow candidate using the $^{13}$CO and HCO$^+$ maps.
The results of the outflow identification is summarized in Table \ref{tab:protostars}.
We note that the two marginally detected candidates might be just contamination of nearby outflow components from FIR 3 and VLA 13.
To confirm that the HOPS64 and HOPS108 are the driving sources of the molecular outflows, higher-angular resolution observations will
be needed. 
The scenario that the dynamical compression due to the  outflows has triggered star cluster formation 
in this region \citep{shimajiri08} is not inconsistent with the chemical timescale of 10$^5$ years expected from the CCS abundance.
Star formation in FIR 4 may have been triggered by the outflow compression.
Alternative scenario is that star formation in FIR 4 happened before the outflow hit the FIR 4 clump which evolved into a hot core \citep{kama13}
and subsequent star formation in FIR 4, or fragmentation,  may be triggered by the outflow compression \citep{shimajiri08}.
Further chemical modeling  may provide us a hint to constraint the evolution of FIR 4.

\section{Summary}
\label{sec:summary}

We summarize the main results of the present paper as
follows.

\begin{itemize}
\item[1.] We carried out mapping observations toward an outflow-shocked region, OMC-2, FIR 3/4/5 
in $^{13}$CO ($J=1-0$), C$^{18}$O ($J=1-0$), N$_2$H$^+$  ($J=1-0$), CCS  ($J_N=8_7-7_6$), HCO$^+$  ($J=1-0$), 
H$^{13}$CO$^+$  ($J=1-0$), HN$^{13}$C  ($J=1-0$), H$^{13}$CN  ($J=1-0$), DNC ($J=1-0$), N$_2$D$^+$ ($J=1-0$), and DC$_3$N ($J=9-8$) using 
a new 4 beam receiver FOREST and the T70 receiver. 
We detected all the molecular lines except N$_2$D$^+$ and DC$_3$N.

\item[2.] We detected faint CCS emission from a dense clump containing FIR 4.
This is the first detection of CCS emission in OMC-2.
The typical fractional abundance is estimated to be about a few $\times 10^{-12}$.


\item[3.]  The [HNC]/[HCN] ratio is smaller than unity in the N$_2$H$^+$ clump
and the outflow lobes, whereas the [HNC]/[HCN] ratio is sometimes higher in the clump envelope. 

\item[4.]  From the N$_2$H$^+$ hyperfine fit, we found that the excitation temperature 
in this region is high. The high excitation temperature might be due to the effect of radiation by protostars in this region.

\item[5.] 
We confirmed 4 known outflow candidates including marginal detection by \citet{takahashi08} and 
 1  new possible outflow candidate using $^{13}$CO and HCO$^+$ ($J=1-0$) emission.
Previous studies suggest that the FIR 3 outflow hits the FIR4 clump \citep{shimajiri08,gonzalez16} and triggered star formation in this region \citep{shimajiri08}.
In the present paper, we proposed that the protostellar outflow from VLA13 as well as FIR3 may have compressed the FIR 4 clump, 
taking into account the spatial distribution of blueshifted and redshifted $^{13}$CO and HCO$^+$ components projected on the plane-of-the sky.
In addition, 
there is a possibility that the moderately-small [HNC]/[HCN] ratios in the N$_2$H$^+$ clump and the outflow lobes
might be due to the dynamical interaction between the clump and outflows and strong FUV radiation, both of
which might release the atomic carbon that prevents the formation of HNC.

\end{itemize}

\begin{ack}
This work was carried out as one of the large projects of the Nobeyama 
Radio Observatory (NRO), which is a branch of the National Astronomical 
Observatory of Japan, National Institute of Natural Sciences. 
We thank the NRO staff for both operating the 45 m and helping us with the data reduction. 
This work was financially supported by Grant-in- Aid for Scientific Research (Nos. 17H02863, 17H01118,
26287030).
YS received support from the ANR (project NIKA2SKY, grant agreement ANR-15-CE31-0017).
\end{ack}

\bibliographystyle{apj}
\bibliography{orionco}

\clearpage

\end{document}